\documentclass[manuscript]{aastex62}

\newcommand{\sm}{$M_\odot$}

\newcommand{\iras}{IRAS 15398$-$3359}

\newcommand{\hhco}{H$_{2}$CO}

\newcommand{\ccchh}{c-C$_3$H$_2$}
\newcommand{\meta}{CH$_3$OH}
\newcommand{\kms}{km s$^{-1}$}
\newcommand{\mjybeam}{mJy beam$^{-1}$}

\received{,}
\revised{, }
\accepted{}
\submitjournal{ApJ}

\shorttitle{}

\begin{document}
\title{Molecular Distributions of the Protostellar Envelope and the Outflow of \iras: Principal Component Analysis}

\correspondingauthor{Yuki Okoda}
\email{}

\author[0000-0002-0786-7307]{YUKI OKODA}

\author{YOKO OYA}
\affiliation{Department of Physics, The University of Tokyo, 7-3-1, Hongo, Bunkyo-ku, Tokyo 113-0033, Japan; okoda@taurus.phys.s.u-tokyo.ac.jp}

\author{NAMI SAKAI}
\affiliation{RIKEN Cluster for Pioneering Research, Wako, Saitama 351-0198, Japan}

\author{YOSHIMASA WATANABE}
\affiliation{Materials Science and Engineering, College of Engineering, Shibaura Institute of Technology, 3-7-5 Toyosu, Koto-ku, Tokyo 135-8548, Japan}



\author{SATOSHI YAMAMOTO}
\affiliation{Department of Physics, The University of Tokyo, 7-3-1, Hongo, Bunkyo-ku, Tokyo 113-0033, Japan; okoda@taurus.phys.s.u-tokyo.ac.jp}



\begin{abstract}
\par We have imaged 15 molecular-line emissions and the dust continuum emission around the Class 0 protostellar source, \iras, with ALMA.
The outflow structure is mainly traced by the \hhco\ ($K_a=0$ and 1), CCH, and CS emissions.
These lines also trace the disk/envelope structure around the protostar.
The \hhco\ ($K_a=2$ and 3), \meta, and SO emissions are concentrated toward the protostar, while the DCN emission is more extended around the protostar.
We have performed the principal component analysis (PCA) for these distributions on the two different scales, the outflow and the disk/envelope structure.
For the latter case, the molecular-line distributions are classified into two groups, according to the contribution of the second principal component, one having a compact distribution around the protostar and the other showing a rather extended distribution over the envelope.
Moreover, the second principal component value tends to increase as an increasing quantum number of \hhco\ ($K_a=0,1,2,$ and 3), reflecting the excitation condition: the distribution is more compact for higher excitation lines.
These results indicate that PCA is effective to extract the characteristic feature of the molecular line distributions around the protostar in an unbiased way.
In addition, we identify four blobs in the outflow structure in the \hhco\ lines, some of which can also be seen in the \meta, CS, CCH, and SO emissions.
The gas temperature derived from the \hhco\ lines ranges from 43 to 63 K, which suggests shocks due to the local impact of the outflow on clumps of the ambient gas.

\end{abstract}



\section{Introduction} \label{sec:intro}

\par Observations of protostellar envelope/disk systems around very young protostars are of fundamental importance in elucidating physical and chemical evolution from protostellar cores to planetary systems.
Thanks to high-resolution and high-sensitivity molecular-line observations with the Atacama Large Millimeter/submillimeter Array (ALMA), a basic kinematic structure around a protostar, that is a rotationally supported disk surrounded by an infalling-rotating envelope, has been revealed for a number of low-mass protostellar sources \citep*[e.g.,][]{Codella et al.(2014), Sakai et al.(2014a), Sakai et al.(2014b), Ohashi et al.(2014), Oya et al.(2014), Yen et al.(2015), Aso et al.(2017), Oya et al.(2016), Oya et al.(2018)}.
Moreover, a complex chemical structure in a disk/envelope system has also been recognized: molecular distributions reflect the physical conditions, and hence, they are different from molecule to molecule and even from transition to transition. \citep*[e.g.,][]{Sakai et al.(2014a), Oya et al.(2016), Oya et al.(2017), Oya et al.(2019), Lee et al.(2017), Jacobsen et al.(2019)}.
It is thus suggested that the molecular distributions can be used as a useful tool to study physical processes of disk formation in the envelope.
\par Recently, sensitive observations with a broad instantaneous bandwidth become popular in various radio telescopes including ALMA, so that many molecular lines can be observed at a single observation. 
This situation enables us to obtain rich information on chemical composition as well as physical structure.
On the other hand, it takes huge efforts to characterize the distributions of all the observed lines in a one-by-one way \citep*[e.g.,][]{Jorgensen et al.(2016), Imai et al.(2016), Watanabe et al.(2017)}. 
For this reason, physical and chemical structures are often explored by using only a few selected lines, where interpretations could be biased by the selection.
To make a full use of the observed lines without any preconception, introduction of a machine learning process is essential. 
\par A principal component analysis (PCA) is a powerful method to characterize the distributions comprehensively \citep{Jolliffe(1986)}.
This method has already been used for radio astronomical observations of molecular lines.
It was conducted for large scale observation data for external galaxies and galactic molecular clouds \citep*[e.g.,][]{Ungerechts et al.(1997), Meier & Turner(2005), Watanabe et al.(2016)}. 
\cite{Spezzano et al.(2017)} used this method to investigate the chemical structure of the L1544 starless core 
and successfully highlighted four characteristic distributions.
Thus, it is now important to apply this method to various sources including protostellar sources.

\par \iras\ is a low-mass Class 0 protostellar source \citep*[$T_{\rm bol}$=44 K;][]{Jorgensen et al.(2013)} in the Lupus 1 molecular cloud \citep*[$d=$155 pc;][]{Lombardi et al.(2008)}. 
Star formation activities of this source were studied by observing a molecular outflow in the CO lines with single-dish telescopes \citep{Tachihara et al.(1996), van Kempen et al.(2009)}. 
On the other hand, chemical characteristics of this source were investigated by \cite{Sakai et al.(2009)}.
Based on single dish observations, they identified this source as a warm carbon-chain chemistry (WCCC) source, which is rich in various carbon-chain related molecules such as CCH, $\rm C_{4}$H, and C$\rm H_3$CCH on a few 1000 au scale around the protostar.
These molecules are produced through the gas phase reactions triggered by evaporation of CH$_4$ from the grain mantles in a lukewarm region (T$\sim$30 K).
\iras\ was actually the second WCCC source discovered after L1527 \citep{Sakai et al.(2008), Sakai&Yamamoto(2013)}.

\par This source has been a good target for star formation studies in the ALMA era.
\cite {Jorgensen et al.(2013)} conducted sub-arcsecond resolution observations toward \iras.
A ring structure of the H$^{13}\rm CO^{+}$ ($J=4-3$) emission was found at a 150$\sim$200 au scale, which was proposed to be a sign of a recent accretion burst: H$^{13}\rm CO^{+}$ seems to be destroyed through the gas-phase reaction with H$_2$O evaporated due to the enhanced luminosity by the episodic accretion.
The ALMA observations of the HDO ($1_{0,1}-0_{0,0}$) emission revealed that it locally resides on the cavity wall near the protostar, which was also interpreted in terms of the recent accretion burst \citep{Bjerkeli et al.(2016b)}.

\par \cite {Oya et al.(2014)} characterized the bipolar outflow extending along the northeast-southwest axis on a 2000 au scale by observing the $\rm H_2$CO ($5_{1,5}-4_{1,4}$) line with ALMA at a resolution of 0\farcs5 ($\sim$80 au).
By analyzing the kinematic structure of the outflow, they derived the inclination angle of the disk/envelope structure to be 70\degr\ (0\degr\ for a face-on configuration), meaning that the outflow is blowing almost parallel to the plane of the sky.
This result is further confirmed by the CO observation with SMA \citep{Bjerkeli et al.(2016a)}. 
The upper limit to the protostellar mass is derived to be 0.09 $M_\odot$ from the velocity structure of the $\rm H_2$CO ($5_{1,5}-4_{1,4}$) emission \citep{Oya et al.(2014)}, and 0.01 $M_\odot$ from the analysis of the $\rm C^{18}$O ($J=2-1$) line \citep{Yen et al.(2017)} at a resolution of 0\farcs5.
In spite of such a low protostellar mass, \cite{Oya et al.(2014)} reported a high velocity component in the \hhco\ emission toward the protostar position, which could possibly be ascribed to a disk structure.
 Indeed, \cite{Okoda et al.(2018)} found the Keplerian disk structure based on the observation of the SO line at a higher angular resolution($\sim$0\farcs2) with ALMA and determined the protostellar mass to be 0.007 $^{+0.004}_{-0.003}$ \sm.
Since this source is deeply embedded in a parent dense gas \citep{Kristensen et al.(2012), Jorgensen et al.(2013)}, the protostar is likely to be young.
Hence, the result by \cite{Okoda et al.(2018)} indicates that the disk formation has already started around a newly born protostar in its very early evolutionary stage.
\par In this protostellar source, various molecular lines are detected on different scales with ALMA.
In this paper, we apply the PCA to the molecular-line and continuum images observed toward this source to characterize their distributions.




\par 
\section{Observation} \label{sec:style}
\par The ALMA observations were carried out toward \iras\ in the Cycle\ 2 and Cycle\ 3 operations.
Major points are summarized below.
\par The Cycle 2 observation was conducted on 2015 July 20. 
Spectral lines of CCH, SO, and CS listed in Table \ref{observations} were observed in the frequency range from 244 to 263 GHz with the Band 6 receiver. 
The original synthesized beam size is 0\farcs21$\times$0\farcs15 (P.A. 58\arcdeg) for the continuum image, while those of the line images are summarized in Table \ref{observations}.
The rms noise levels for the continuum emission, the CCH emission, the SO emission, and the CS emission are 0.12 mJy bea$\rm m^{-1}$, 4 mJy bea$\rm m^{-1}$, 4 mJy bea$\rm m^{-1}$, and 4 mJy bea$\rm m^{-1}$, respectively, at a resolution of 61 kHz.
Other details of the observations are described elsewhere \citep{Okoda et al.(2018)}.

\par The Cycle 3 observation was conducted on 2016 March 31.
Spectral lines of \hhco, \ccchh, CCH, \meta, and DCN listed in Table \ref{observations} were observed in the frequency range from 349 to 365 GHz with the Band 7 receiver. 
Forty-two antennas were used in the observations, where the baseline length ranged from 14.70 to 452.72 m. 
The field center was  ($\alpha_{2000}$, $\delta_{2000}$)= (15\fh43\fm02\fs242, $-$34\arcdeg 09\arcmin 06\farcs70), which was the same as that of the Cycle 2 observation.
The total on-source time was 19.30 minutes.
The primary beam (half-power beam) width was 17\farcs08.
The backend correlator for molecular line observations except for the DCN observation was set to a resolution of 122 kHz and a bandwidth of 59 MHz, and that for the DCN observation was set to a resolution of 977 kHz and a bandwidth of 938 MHz.
It should be noted that the velocity of all spectral windows is blue-shifted by 2.5 \kms\ due to the fault in the observation setting.
Hence, we do not discuss the velocity structures and only focus on the distribution of the molecules. 
The original synthesized beam size of the continuum image is 0\farcs48$\times$0\farcs45 (P.A. 83\arcdeg), while those of the lines are summarized in Table \ref{observations}.
\par Images were prepared by using the CLEAN algorithm, where the Briggs' weighting with a robustness parameter of 0.5 was employed. 
The continuum image was obtained by averaging line-free channels, and the line images were obtained after subtracting the continuum component directly from the visibilities. 
Self-calibration was not applied in this study, because the continuum emission is not bright enough.
Since these largest angular sizes are 2\farcs5 for both of these observations, the intensities of the structures extended more than that size could be resolved-out.
\par In order to compare the molecular distributions at the same spatial resolution, the beam size was set to be  0.5$''\times0.5''$ by using a gaussian kernel with $imsmooth$, which is the task of the Common Astronomy Software Applications package (CASA) \citep{McMullin et al.(2007)} to smooth a data cube across spatial dimensions.
The pixel sizes were set to be (2048, 2048) for the whole observed area in the CLEAN procedure. 
\par
\begin{figure}
\centering
\includegraphics[scale=0.7]{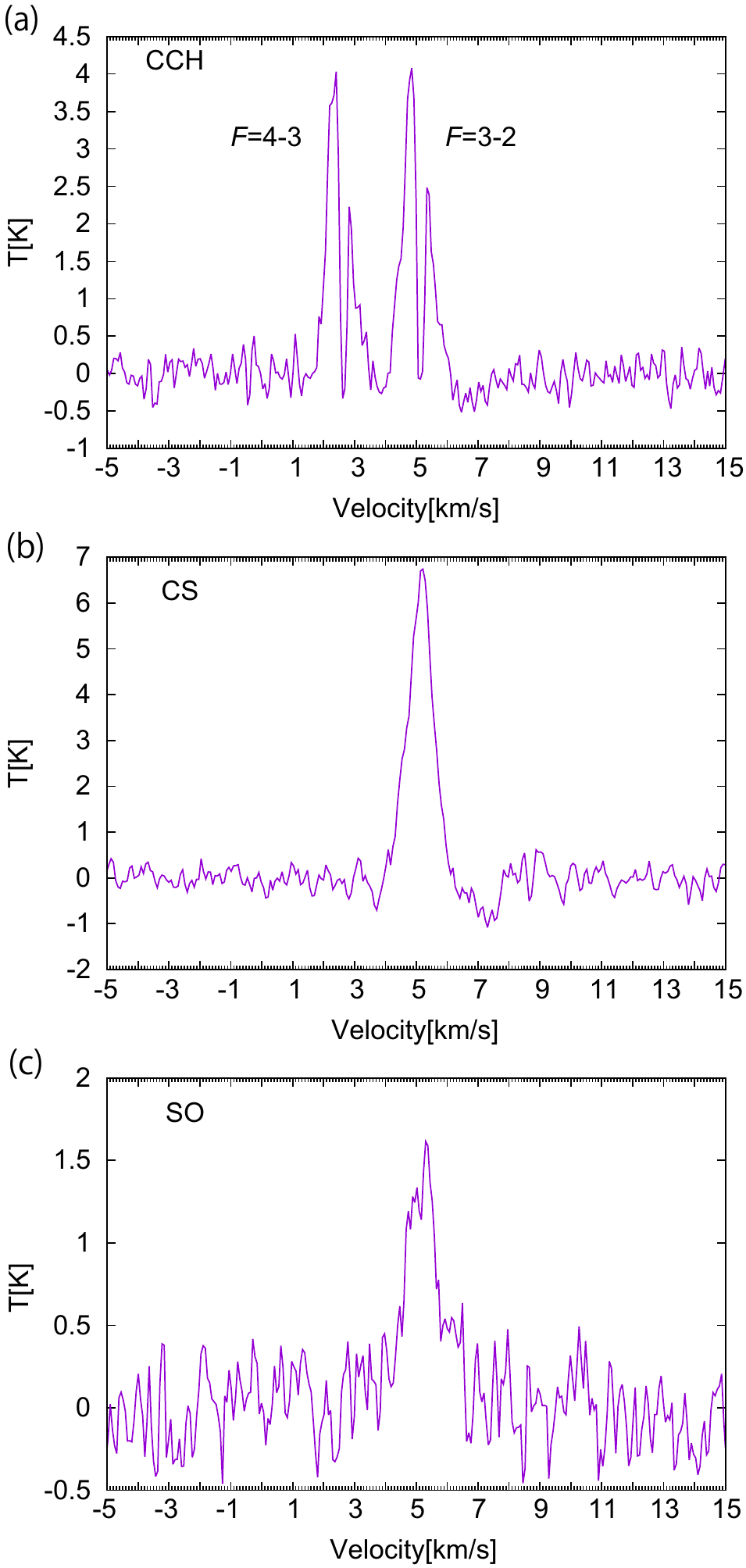}
\caption{Examples of the spectra of the CCH ($N=3-2, J=7/2-5/2, F=4-3$ and $3-2$), CS, and SO lines observed toward the protostellar position. A systemic velocity is 5.2 \kms. In the CCH spectrum, two hyperfine components ($F=4-3$ and $3-2$) are seen, where an intensity dip near the center of each component is due to self-absorption by a foreground gas. In the imaging, these two lines are stacked to improve the signal-to-noise ratio (See Section 2).\label{spectra}}
\end{figure}

 Now, we have 16 images including the continuum emission.
It should be noted that there are a few pairs of lines unresolved in the observation.
The velocity width of the line is typically 1.0$-$1.5 \kms, as shown in Figure \ref{spectra}, which corresponds to 0.9$-$1.3 MHz and 1.2$-$1.8 MHz for the Band 6 and Band 7 observations, respectively.
Hence, the hyperfine splitting of CCH ($F=5-4$ and $4-3$ of the $N=4-3, J=9/2-7/2$ transition) and that of CCH ($F=4-3$ and $3-2$ of the $N=4-3, J=7/2-5/2$ transition) are not resolved.
Similarly, the 9$_{1,8}-8_{2,7}$ and 9$_{2,8}-8_{1,7}$ lines of \ccchh\ are degenerated, and the 10$_{0,10}-9_{1,9}$ and 10$_{1,10}-9_{0,9}$ lines of that are, too.
These unresolved pairs were treated as a single line.
For the CCH ($N=3-2$) lines ($F=4-3$ and $3-2$), a single image was prepared in order to increase the signal-to-noise ratio.


\section{Molecular Distributions} \label{sec:style}
\par Figure \ref{outflow} shows the continuum map and the moment 0 maps of the observed molecular lines, while Figure \ref{protostar} depicts their blow-ups around the protostar.
Figures \ref{outflow}(a) and \ref{protostar}(a) show the 0.8 mm continuum distribution (Cycle 3), whose peak intensity is 23.95$\pm$0.48\mjybeam.
The coordinates of the peak are derived from a 2D Gaussian fit to the image: ($\alpha_{2000}$, $\delta_{2000}$)= ($15\fh43\fm02\fs2359\pm0.0004, $-$34\arcdeg 09\arcmin 06\farcs8348\pm0.0045$), which are consistent with the previous reports \citep*[e.g.,][]{Oya et al.(2014), Okoda et al.(2018)}.
The continuum emission has a single peak with a circular distribution.
\par As shown in Figure \ref{outflow}, the distribution is different from molecular line to molecular line.
The outflow structure along the northeast-southwest axis reported previously \citep{Oya et al.(2014), Bjerkeli et al.(2016a), Okoda et al.(2018)} can be seen particularly in the CCH, CS, and \hhco\ ($K_a=0$ and 1) lines.
The outflow seems to have the double ring structures in the CS emission.
This may be caused by the episodic accretion \citep{Bjerkeli et al.(2016a)}. 
The \ccchh\ emission does not have a component clearly associated with the protostar. 
It traces a part of the outflow cavity wall of the southwestern side.
DCN seems to be distributed in the outflow cavity wall to some extent and also has a compact distribution around the protostar.
The SO, \meta, and \hhco\ ($K_a=2$ and 3) emissions reveal blobs in the outflow as well as the compact distribution around the protostar.
\par In the blow-up version of the moment 0 maps (Figure \ref{protostar}), the component associated with the protostar can be found in most of the observed lines except for the \ccchh\ lines.
The disk/envelope structure perpendicular to the outflow direction is seen more clearly.
The CCH emission traces the envelope extending from northwest to southeast \citep{Okoda et al.(2018)}.
Note that the distribution of the CCH ($N=4-3, J=7/2-5/2, F=3-3$) line looks slightly different from those of the other CCH lines.
This is probably due to the low signal-to-noise ratio of the line.
The CS emission seems to be distributed over the envelope around the protostar.
However, the CS emission in the southeastern side of the envelope is brighter than that in the northwestern side, as shown in Figure \ref{protostar}(b).
For the CS emission, there is asymmetry in the distribution around the protostar, which is also seen for the \hhco\ ($K_a=$0) emission.
As above, the distributions of the observed molecular emissions look different from one another.
However, their classification by eye may suffer from our preconception, and hence, an objective way to characterize the distributions is needed.


\section{Principal Component Analysis}
\par We conduct a principal component analysis (PCA) \citep{Jolliffe(1986)} for the two scales shown in Figures \ref{outflow} and \ref{protostar} to explore similarities and differences of molecular line distributions and the Cycle 3 continuum distribution in an unbiased way.
We exclude the Cycle 2 continuum data from the analysis, because the signal-to-noise ratio is lower than that of the Cycle 3 continuum data.
Thus, we use 16 dimension dataset which consists of 15 molecular line data and the Cycle 3 continum data for the whole structure.
For the disk/envelope structure, we exclude \ccchh\ (9$_{1,8}-8_{2,7}$ and 9$_{2,8}-8_{1,7}$) because there are not enough data above the threshold level defined below.
We write the code for the PCA by using python libraries, $numpy, matplotlib, pandas$, and $astropy$.
\par We represent the observed distribution of the $j$ th emission as the vector {\bf x}$_{j}$.
In PCA, we look for the functions {\bf z}$_i$ ($i$=1-16 or 1-15) formed by a linear combination of data {\bf x}$_{j}$ ($j$=1-16 or 1-15), where the $i$ th function is uncorrelated with all the others.
The main features can be extracted from the multidimensional data set by picking up a few components of {\bf z}, which have a large variance. 
Thus, this process means "reduction" of the dimension.
In this paper, these functions are obtained by diagonalizing the correlation matrix, because the intensities are different from molecular line to molecular line.
In the calculation of the correlation coefficients between two intensity distributions, the data above threshold levels for the both distributions are used.
Three times the rms noise level (Tables \ref{corr_outflow}) is employed as the threshold level.
The correlation matrices for the 16 or 15 distributions (15 or 14 lines and one continuum) thus obtained for the whole structure and the disk/envelope structure are shown in Tables \ref{corr_outflow} and \ref{corr_protostar}, respectively.
The matrix is diagonalized to find the orthogonal linear functions called as the principal components.
Here, their eigenvalues are also derived at the same time.
The $i$ th principal component {\bf z}$_i$ is denoted as PC$i$ ($i$=1-16 or 1-15), where the number $i$ is assigned in a decreasing order of the eigenvalue (Tables \ref{vec_outflow} and \ref{vec_protostar} for the whole structure and the disk/envelope structure, respectively).
As the correlation matrix is used to find a set of orthogonal linear functions, the principal components are defined as:
\begin{equation}
{\bf z}={\bf Ax^*},
\end{equation}
and
\begin{equation}
{\bf x^*}_j={\bf x}_j/\sigma_{jj}^{1/2},
\end{equation}
where {\bf A} is the transformation matrix for the diagonalization of the correlation matrix, {\bf x*} the normalized distribution, and $\sigma_{jj}$ the variance of  {\bf x}$_{j}$.
Hence, PC$i$ is dimensionless. 
A contribution ratio of each principal component is caluculated by dividing the eigenvalue by the dimension (16 or 15), as shown in Tables \ref{vec_outflow} and \ref{vec_protostar}.
It represents how much PC$i$ contributes to all the original sample distributions.

\subsection{PCA of the whole structure}
\par As mentioned above, we can almost reproduce the molecular-line and continuum distributions by using only a few principal components.
According to Table \ref{vec_outflow}, PC1 and PC2 have the two largest contribution ratios, 42.1 \% and 20.3 \%, respectively.
The sum of the contribution ratios of the two components is 62.4 \%, which indicates that these two components can be regarded as the main components.
The contribution ratios of PC3 and PC4 are 8.7 \% and 8.2 \%, respectively, which are significantly smaller than that of PC2.
Hence, we here discuss the first two components.
In this case, the observed images can approximately be reproduced by the linear combinations of the two component images, which means the reduction of the original dimension of the images (16) to 2.
PC3 and PC4 are discussed in Appendix A for reference.

Figures \ref{pc_outflow}(a) and \ref{pc_outflow}(b) show the maps of PC1 and PC2, respectively.
PC1 represents an overall shape of the outflow, two blobs in the outflow, and a component concentrated around the protostellar position.
PC2 mainly represents the disk/envelope structure.
The foot of the outflow can also be seen partly.
\par Contributions of the first two principal components for each molecular-line distribution are represented on the PC1-PC2 plane, as shown in Figures \ref{pc_outflow}(c).
Correlations between PC $i$ ($i=$1$-$4) and the observed distributions are also presented in Appendix B.
Grey dashed ellipses in Figure \ref{pc_outflow}(c) represent the estimated errors due to noise, whose details are described in Section 4.3.
The majority of the line emissions and the continuum emission show the positive values of PC1, as shown in Figure \ref{pc_outflow}(c).
For this case, the positive and negative values on the PC2 axis can classify the samples into two groups, one with a compact distributions around the protostar (Group 1) and the other showing rather extended structures (Group 2), respectively.
The CCH and CS lines as well as the \hhco\ ($K_a=0$) line belong to Group 2, which well trace the extended feature of the outflow. 
Indeed, these lines are often employed as a tracer of the outflow cavity wall on a 1000 au scale \citep*[e.g.,][]{Codella et al.(2014), Oya et al.(2015), Oya et al.(2019), Zhang et al.(2018)}.
On the other hand, \ccchh\ (10$_{0,10}-9_{1,9}$ and $10_{1,10}-9_{0,9}$), \ccchh\ (9$_{1,8}-8_{2,7}$ and 9$_{2,8}-8_{1,7}$) and CCH ($N=4-3$, $J=7/2-5/2$, $F=3-3$) have negative PC1 and positive PC2.
This result represents that these molecular lines have a clumpy feature in the outflow.
\par The molecular emissions distributed in the blobs are represented by the positive PC1 and positive PC3 values (see Appendix A).
These blobs are most likely a shocked region caused by a local impact of the outflow on an ambient gas, as described below.
The molecular lines showing the positive PC1 and the positive PC3 (\hhco, CS, CCH ($N=3-2$), SO, and \meta) trace both or one of the two blobs (Figure \ref{outflow_pc3_4}(c)).
We finally identify four blobs (A-D in Figure \ref{temperature_region}) in the outflow structure by using the \hhco\ ($K_a=$1) line (Table \ref{position}).
They are also seen in the other \hhco\ lines, although blobs B and C can hardly be seen in the highest excitation lines of \hhco\ ($K_a=3$).
Blob D is bright in the CS and SO emissions, although blob A is seen in the CS and CCH ($N=3-2$) lines.
The \meta\ emission has an extended distribution from the protostar to blob A.
\par The gas kinematic temperature for each blob is evaluated from the detected \hhco\ lines under the non-LTE (local thermodynamic equilibrium) method assuming the large velocity gradient (LVG) approximation.
The data used in the analysis are shown in Table \ref{lines}.
The derived gas kinematic temperature ranges from 43 K to 63 K (Table \ref{tempvalue}).
Such high temperatures as well as absence of associated continuum emission indicate that the four blobs should be the shocked regions caused by the outflow impact.
It seems that a local shock is occurring on the cavity wall by the interaction with an ambient gas.
This situation is also pointed out for blob A by \cite{Oya et al.(2014)}.
Such a shocked region can be seen in L1157 B1, where strong emissions of various molecules including \hhco, CS, SO, and \meta\ are detected \citep*[e.g.,][]{Bachiller(1997), Benedettini et al.(2007), Codella et al.(2010)}.
These molecules are thought to be liberated from dust grains and/or produced through the gas-phase shock chemistry.
Detailed comparison with the L1157 B1 result for exploring shock chemistry would require observations of more molecular lines including SiO.

\subsection{PCA of the disk/envelope structure}
\par The PCA for the observed distributions in the narrower range (Figure \ref{protostar}) is helpful to investigate the chemical structure around the protostar  in more detail.
We use 14 molecular line data except for \ccchh\ (9$_{1,8}-8_{2,7}$ and 9$_{2,8}-8_{1,7}$) in addition to the Cycle 3 continuum data.
Table \ref{vec_protostar} shows the eigenvalues and eigenvectors obtained by diagonalizing the correlation matrix.
On the disk/envelope scale, PC1 and PC2 stand for 76.7 \% of the contribution ratio. 
The molecular distributions can mostly be reproduced by only the first two components, so that we here discuss PC1 and PC2.
The contribution ratio of PC3 is 8.5 \%, which is similar to that of PC4 (7.8 \%).
PC3 and PC4 are discussed in Appendix A for reference.
Correlation between PC $i$ ($i=$1$-$4) and the observed distributions are presented in Appendix B.
\par In Figure \ref{pc_protostar}(a), PC1 shows the distribution centered near the protostellar position with extension along the northwest-southeast direction.
This component apparently represents the disk/envelope structure.
All the molecular lines except for CCH ($N=4-3, J=7/2-5/2, F=3-3$) and \ccchh\ (10$_{0,10}-9_{1,9}$ and $10_{1,10}-9_{0,9}$) have the positive PC1 component, as shown in Figures \ref{pc_protostar}(c).
Here, grey dashed ellipses represent the estimated errors (Section 4.3).
The exception for CCH ($N=4-3, J=7/2-5/2, F=3-3$) and \ccchh\ (10$_{0,10}-9_{1,9}$ and $10_{1,10}-9_{0,9}$) means that they are not mainly distributed in the disk/envelope system, as noted in Section 3 and 4.1.
PC2 has negative values at almost all the positions with two large negative peaks at the both sides of the protostellar position (Figure \ref{pc_protostar}(b)).
Its 'red part' in the northeastern and southwestern extension is close to zero.
SO shows the large positive value for the PC2 axis, as shown in Figure \ref{pc_protostar}(c).
This result means that its distribution is very compact toward the protostar: the southeastern and northwestern extension of PC1 is almost compensated by the positive PC2.
It is consistent with our previous finding of the compact SO distribution \citep{Okoda et al.(2018)}.
Similarly, the lines showing positive PC1 and positive PC2 have a compact distribution around the protostar (Group A).
On the other hand, the lines showing positive PC1 and negative PC2 have a rather extended distribution (Group B).
Thus, PC2 can be an indicator of how much the distribution is concentrated.
\par For the two large negative peaks of PC2, the southeastern side of the protostar has a stronger peak than the northwestern side (Figure \ref{pc_protostar}(b)).
This feature contributes to an asymmetric distribution in the disk/envelope system.
The CS and \hhco\ ($K_a=0$) distributions are clearly brighter in the southeastern side, and hence, the PC2 values take a negative value.
The chemical composition seems azimuthally non-uniform even in the protostellar envelope, as revealed in the other sources \citep{Sakai et al.(2016), Oya et al.(2017)}.
\par Moreover, we focus on the behavior of the \hhco\ lines with different upper state energies along the PC2 axis in Figure \ref{pc_protostar}(c).
For para \hhco, the $K_a=2$ line shows the positive contribution of PC2, while the $K_a=0$  line shows the negative contribution.
Likewise, ortho \hhco\ reveals a similar trend: the $K_a=3$ lines take the larger positive value of PC2, while the $K_a=1$ line takes the negative PC2.
This trend represents that the emissions of the higher excitation lines (i.e., higher $K_a$ lines) tend to be more concentrated around the protostar.
This is reasonable because the gas temperature and the gas density are expected to be higher as approaching to the protostar.
The gas kinematic temperature is calculated to be 55 K (Table \ref{tempvalue}) by using the non-LTE calculation applied to the analysis of the blobs.
The gas is significantly heated near the protostar.

\subsection{Effect of noise}
\par In order to evaluate the uncertainties for the PCA \citep{Gratier et al.(2017), Spezzano et al.(2017)},  we derive the standard deviations for the principal component values of each molecular line in the following way. 
We generate Gaussian random noise for each pixel of the image field of each molecular or continuum distribution, and the noise distribution is convolved by the beam size of the observation (0.5$''\times$0.5'').
The standard deviation of the convolved noise image is adjusted to be one standard deviation of the observed noise.
The artificial noise thus prepared for each molecular or continuum distribution is added to the observed distribution, and then, the PCA is conducted.
This procedure is repeated 1000 times, and the standard deviations of the PCA components are finally calculated.
\par In Figures \ref{pc_outflow}(c), \ref{pc_protostar}(c), \ref{outflow_pc3_4}(c,d), and \ref{protostar_pc3_4}(c,d), the grey dashed ellipses represent the uncertainties, whose major and minor axes are the standard deviation for each axis.
 On the whole-structure scale, all the standard deviation for PC1 is lower than 0.1.
\ccchh\ (10$_{0,10}-9_{1,9}$ and $10_{1,10}-9_{0,9}$) shows the largest standard deviation for PC2 (0.17).
The standard deviation for most molecular lines is less than 0.1 for PC2.
Since the signal-to-noise ratio is high for the \hhco\ ($K_a=$0) line, the standard deviation is smaller than 0.06 for PC1 and PC2.
For PC3 and PC4, the standard deviation is from 0.06 to 0.2, which is slightly larger than those for PC1 and PC2 (Appendix A).
\par The standard deviation of the PCA components for the disk/envelope scale is comparable to that for the whole-structure scale.
Even the molecular lines having the rather poor signal-to-noise ratio, such as \ccchh\ (10$_{0,10}-9_{1,9}$ and $10_{1,10}-9_{0,9}$), CCH ($N=4-3, J=7/2-5/2, F=3-3$), \meta, and DCN, show the standard deviation smaller than 0.1 for PC1.
\ccchh\ (10$_{0,10}-9_{1,9}$ and $10_{1,10}-9_{0,9}$), CCH ($N=4-3, J=7/2-5/2, F=3-3$), and DCN have the standard deviation smaller than 0.15 for PC2.
While most lines have the standard deviation of 0.1$-$0.2 for PC3 and PC4, CCH ($N=4-3, J=9/2-7/2, F=4-3$), ($N=4-3, J=7/2-5/2, F=4-3$ and $F=3-2$), \hhco\ ($K_a=$0, 1, and 2), and the dust continuum show the value smaller than 0.1.
As a result, we find that our results of the PCA and the related discussions described above are not essencially changed by the effect of noise.

\section{Comparison with the other protostellar sources}
\par As mentioned in Introduction, \iras\ is regarded as a WCCC source which is rich in carbon-chain molecules on a few 1000 au scale.
Molecular distributions observed in the disk/envelope region ($\sim$100 au scale) of this source are therefore compared with those of the other WCCC sources.
In the prototypical WCCC source L1527, \cite{Sakai et al.(2014a), Sakai et al.(2014b)} reported the chemical structure around the protostar with ALMA.
According to their result, CCH, CS, and \ccchh\ mainly trace the infalling-rotating envelope gas outward of its centrifugal barrier, while SO mainly exists near the centrifugal barrier and partly inward of it.
\hhco\ resides over the disk/envelope region, and \meta\ seems to exist around the centrifugal barrier and in the disk region.
For TMC-1A, which is the WCCC source in the Class I stage, the distributions of CS, SO, and SO$_2$ were observed with ALMA by \cite{Sakai et al.(2016)}.
In this source, CS also traces the infalling-rotating envelope gas, while SO seems to trace the centrifugal barrier.
\par These characteristic features are indeed found in \iras.
In the PCA for the disk/envelope structure, PC2 shows that CCH and CS can be classified to one group (Group B in Figure \ref{pc_protostar}) showing the existence in the infalling-rotating envelope, while SO and \meta\ can be classified to another group (Group A in Figure \ref{pc_protostar}) revealing more compact distributions.
On the other hand, the \hhco\ lines take different PC2 component values depending on their upper-state energy.
This feature of \hhco\ is consistent with that found in L1527, where \hhco\ resides over the disk/envelope region.
\par An exception is \ccchh.
While this species clearly traces the infalling-rotating envelope in L1527, it shows a rather different distribution in \iras.
In order to compare the \ccchh\ abundances between L1527 and \iras,  we derive the column density ratios of \ccchh\ relative to \hhco\ at the intensity peak positions of the \ccchh\ (10$_{0,10}-9_{1,9}$ and $10_{1,10}-9_{0,9}$) line (Figure \ref{protostar}(f)) by using the RADEX code \citep{van der Tak et al.(2007)}. 
Here, the northwestern and southeastern peak positions are (15\fh43\fm02\fs20, $-$34\arcdeg09\arcmin06\farcs80) and (15\fh43\fm02\fs28, $-$34\arcdeg09\arcmin07\farcs40), respectively (Figure \ref{protostar}(f)).
The assumptions are as follows: the ortho para ratio of \hhco\ is 3 (statistical value), the H$_2$ density 10$^6$ cm$^{-3}$, and the gas kinematic temperature from 20 K to 40 K.
We employ only the \hhco\ ($K_a=0$) line to estimate the column density of \hhco, because the other lines ($K_a=1, 2,$ and 3) are weak at the intensity peaks of the \ccchh\ (10$_{0,10}-9_{1,9}$ and $10_{1,10}-9_{0,9}$) line.
On the above assumptions, the column densities of ortho \ccchh\ and ortho \hhco\ for the northwestern side are (0.14$-$0.85)$\times$10$^{14}$ cm$^{-2}$ and (0.72$-$2.5)$\times$10$^{14}$ cm$^{-2}$, respectively.
Those for the southeastern side are (0.12$-$0.75)$\times$10$^{14}$ cm$^{-2}$ and (0.92$-$3.5)$\times$10$^{14}$ cm$^{-2}$, respectively.
A large column density range is due to the assumed temperature range.
The \ccchh/\hhco\ ratio is calculated from the column densities derived at the same assumed temperature.
In this case, the temperature dependence is mitigated.
The ratios for the northwestern and southeastern sides are from 0.2 to 0.3 and from 0.1 to 0.2, respectively.
These values are comparable to that found toward L1527 \citep*[0.17$-$0.36;][]{Sakai et al.(2014b)}.
Hence, the abundance of \ccchh\ is not very different between this source and L1527.
The peculiar distribution of \ccchh\ might be due to the overwhelming contribution of the outflow in this source, which is not significant in L1527.
\par It is interesting to note that the above characteristic distributions of molecules can also be seen in the hybrid sources, L483 and B335, where the envelope and its inner most part show WCCC and  hot corino chemistry, respectively \citep{Imai et al.(2016), Oya et al.(2017)}.
Hot corino chemistry is characterized by rich existence of saturated complex organic molecules such as HCOOCH$_3$ and (CH$_3$)$_2$O \citep{Bottinelli et al.(2004), Sakai&Yamamoto(2013)}.
In these sources, the CCH distribution is extended over the envelope with deficiency toward the protostar.
CS traces the infalling-rotating envelope and the inward component, while SO mainly traces the region within the centrifugal barrier.
Although the central concentration of CS looks different from the case of the WCCC sources including \iras, the overall feature is similar.
At present, the origin of the different feature of CS between the WCCC sources and the hybrid sources is puzzling.
This may originate from an insufficient resolution.
In addition, chemical behavior of sulfur-bearing species in the protostellar core has not been investigated well by the chemical model yet \citep*[e.g.,][]{Aikawa et al.(2008), Aikawa et al.(2012)}.
This will be an important target for future astrochemical study.

\section{Summary}
\par We have imaged \iras\ in 15 molecular lines and the dust continuum emission and have conducted the PCA for characterization of their distributions.
The PCA has been performed on the two different scales.
\par On the whole structure scale, we apply the PCA to 16 dataset consisting of 15 molecular line data and the Cycle 3 continuum data.
PC2 can classify the samples having the positive PC1 into two groups, one with compact distributions around the protostar (Group 1) and the other showing rather extended structures (Group 2) (Figure \ref{pc_outflow}).
The molecular lines in the latter group well trace the outflow structure.
The local peaks in the map of PC1 represent the blobs in the outflow, which are probably shocked regions caused by the impact of the outflow on an ambient gas.
\par On the disk/envelope scale, we use 15 dataset except for \ccchh\ (9$_{1,8}-8_{2,7}$ and 9$_{2,8}-8_{1,7}$) for the PCA.
The combination of PC1 and PC2 shows how much the distribution is concentrated toward the protostar.
More compact distributions of the \hhco\ lines with higher upper-state energies are revealed by PC2.
The characteristic molecular distributions revealed by the PCA are consistent with these in the other WCCC sources, L1527 and TMC-1A. 
Thus, the PCA helps us to characterize the molecular line distributions without preconception.
It should also be noted that the result of the PCA depends on the field range which we select.
Hence, it is important to determine a suitable range of interest for the PCA depending on a scientific purpose.
Expanding the dataset will further improve the analysis of this source.
More systematic observations of various molecular lines toward this source as well as application of the PCA to various sources are awaited.
\\
\par 
The authors are grateful to Jes K. J\o rgensen and Ewine. F. van Dishoeck for useful discussions.
We also thank the reviewer of this paper for invaluable comments. 
This paper makes use of the following ALMA data set:
ADS/JAO.ALMA\#2013.1.01157.S and \#2015.1.01380.S (PI: Yoko Oya). ALMA is a partnership of the ESO (representing its member states), the NSF (USA) and NINS (Japan), together with the NRC (Canada) and the NSC and ASIAA (Taiwan), in cooperation with the Republic of Chile.
The Joint ALMA Observatory is operated by the ESO, the AUI/
NRAO, and the NAOJ. 
The authors thank to the ALMA staff for their excellent support.
This study is supported by Grant-in-Aid
from the Ministry of Education, Culture, Sports, Science, and
Technologies of Japan (25108005, 19H05069 and 19K14753).
Yuki Okoda thanks the Advanced
Leading Graduate Course for Photon Science (ALPS) and Japan Society for the Promotion of Science (JSPS)
for financial support.

\appendix
\section{The third and fourth Principal Components}
In the PCA, the first two principal components (PC1 and PC2) can almost reproduce the molecular distributions on the both scales.
PC3 and PC4 have the similar contribution ratios, which are significantly smaller than PC2. 
Nevertheless, PC3 and PC4 show some trends in the distributions.
Here, we briefly discuss their features.
\par On the whole structure scale, PC3 and PC4 indicate the characteristic features for a few molecular lines.
Figures \ref{outflow_pc3_4}(a) and (b) show the maps of PC3 and PC4, respectively.
PC3 represents the two blobs in the outflow as PC1, where the southwestern one is brighter in PC3 than the northeastern one in contrast to PC1.
Hence, the SO line showing a southwestern blob takes the large PC3 value, as shown in the PC1-PC3 plane (Figure \ref{outflow_pc3_4}(c)).
PC3 also has a negative component near the protostar position, which results in the large positive contribution of \ccchh\ (9$_{1,8}-8_{2,7}$ and 9$_{2,8}-8_{1,7}$) and the negative contribution of DCN  (Figure \ref{outflow_pc3_4}(c)).
Meaning of PC4 is not as clear as PC3.
According to the correlations between PC4 and the molecular distributions (Figure \ref{load}), PC4 mainly contributes to representing the peculiar distribution of \ccchh\ (10$_{0,10}-9_{1,9}$ and $10_{1,10}-9_{0,9}$).
As shown in the PC1-PC4 plane, \ccchh\ (10$_{0,10}-9_{1,9}$ and $10_{1,10}-9_{0,9}$) has the large negative value along PC4 axis.
\par As in the case of the whole structure scale, PC3 and PC4 also represent the peculiar distributions of some molecular lines on the disk/envelope scale.
PC3 has the positive and negative distributions around the protostar position in Figure \ref{protostar_pc3_4}(a).
Hence, it represents the asymmetry of the distribution in the disk/envelope structure.
According to the correlations between PC3 and the molecular distributions (Figure \ref{load}), PC3 represents a characteristic feature of DCN and \hhco\ ($K=3_u$): these molecular lines have the large PC3 value (Figure \ref{protostar_pc3_4} (c)).
PC4 solely looks like the distribution of CCH ($N=3-2, J=7/2-5/2, F=4-3$ and $3-2$) (Figures \ref{protostar_pc3_4} (b) and \ref{protostar} (p)).
It can also be seen in Figure \ref{load}.

\section{Correlation Coefficients of the Principal components to the molecular distributions}
\par We calculate the correlation coefficients between the principal components and the molecular distributions on the whole structure and disk/envelope scales, as shown in Figures \ref{load}(a) and \ref{load}(b), respectively. 
These results help us to understand which molecular line distributions the principal component contributes to.
These values are defined as \citep{Jolliffe(1986)}:
\begin{equation}
Cor ({\bf x^*}_j, {\bf y}_i)={\sqrt{\lambda_i}z_{ji}},
\end{equation}
where $\lambda_i$ is the eigenvalue for the $i$ th eigenvector {\bf z$_{i}$}, {\bf y$_{i}$} the distribution for the $i$ th principal component, and $z_{ji}$ the eigenvector component for the $j$ th emission.
On the whole structure scale, the correlation coefficients of PC1 to most of the molecular distributions are higher than 0.5, which means that PC1 represents the overall structure.
PC2 is positively correlated with the molecular distributions mainly showing the clumpy structures in the outflow as discussed in Section 4.1.
PC3 and PC4 are correlated with the peculiar distributions traced by some molecular lines, as described in Appendix A. 
On the disk/envelope scale, PC1 is well correlated with most of the distributions, indicating that it represents the overall disk/envelope structure.
PC2 has a negative correlation with the molecules having the distribution mainly in the envelope around the protostar.
The positive correlation coefficient between PC2 and SO means the compact distribution.
PC3 shows the large correlation with DCN and \hhco\ ($K_a=$3${_u}$).
PC4 is well correlated with CCH ($N=3-2, J=7/2-5/2, F=4-3$ and $3-2$), whose correlation coefficient is almost 1.
This indicates that PC4 represents the peculiar distribution of this line.



\begin{longrotatetable}
\begin{table}[ht]
\centering
\caption{Parameters of Observed Lines $^a$ \label{observations}}
\scalebox{0.7}{
\begin{tabular}{cccccccc}
\hline \hline
 Band $^b$ & Molecule&Transition & Frequency\ (GHz) & $S \mu^2$($D^2$) & $E_{\rm u}$$k^{-1}(\rm K)$ & Original beam size&rms(\mjybeam )\\
 \hline
  6 & CCH&$N=3-2, J=7/2-5/2, F=4-3$& 262.0042600 & 2.3
 & 25 & 0.$''$22 $\times$ 0.$''$16 (P.A. 60$^{\circ}$)&4 \\
 && $N=3-2, J=7/2-5/2, F=3-2$ & 262.0064820 & 1.7
 & 25 &\\ 

6 & SO & $6_{7}-5_{6}$ & 261.8437210 & 16.4
& 47 & 0.$''$22 $\times$ 0.$''$16 (P.A. 55$^{\circ}$)&4\\
 6 &CS &$5-4$ &244.9355565& 19.2& 35 & 0.$''$24 $\times$ 0.$''$17 (P.A. 60$^{\circ}$)&4\\
 
7& \hhco &5$_{0,5}-4_{0,4}$ & 362.7360480 & 27.2 & 52 &0.$''48\times0.''46$ (P.A. 72$^{\circ}$)&9 \\
&&5$_{1,5}-4_{1,4}$& 351.7686450 & 78.3 & 62 & 0.$''50\times0.''47$ (P.A. 78$^{\circ}$) & 10\\
&& 5$_{2,4}-4_{2,3}$&  363.9458940 & 22.8 & 100 & 0.$''48\times0.''45$ (P.A. 77$^{\circ}$)& 10\\
&& 5$_{3,2}-4_{3,1}$& 364.2888840 & 52.2 & 158 & 0.$''48\times0.''45$ (P.A. 78$^{\circ}$) &9\\
&& 5$_{3,3}-4_{3,2}$& 364.2751410 & 52.2 & 158 & 0.$''48\times0.''45$ (P.A. 78$^{\circ}$) &8\\
7&\ccchh &9$_{1,8}-8_{2,7}$& 351.9659690 & 238 & 93 & 0.$''50\times0.47$ (P.A. 79$^{\circ}$) &8\\
&&9$_{2,8}-8_{1,7}$ & 351.9659690 & 79.3 &93 & \\
& &10$_{0,10}-9_{1,9}$& 351.7815780& 101& 96 & 0.$''50\times0.47$ (P.A. 78$^{\circ}$) &8\\
& &10$_{1,10}-9_{0,9}$& 351.7815780& 303 & 96 & \\
7&CCH &$N=4-3, J=9/2-7/2, F=5-4$ & 349.3377056 &2.90 & 42 & 0$''50\times0.''47$ (P.A. 78$^{\circ}$) &9\\
&&$N=4-3, J=9/2-7/2, F=4-3$ & 349.3389882 & 2.32 & 42 \\
&&$N=4-3, J=7/2-5/2, F=4-3$ & 349.3992756 & 2.27 & 42  & 0$''50\times0.''47$ (P.A. 78$^{\circ}$) &9\\
&&$N=4-3, J=7/2-5/2, F=3-2$ & 349.4006712   & 1.69 & 42 \\
&&$N=4-3, J=7/2-5/2, F=3-3$ & 349.4146425 & 0.099 & 42  & 0$''50\times0.''47$ (P.A. 78$^{\circ}$) &8\\
7&\meta &4$_{0,4}-3_{1,3}$ E&  350.6876620 &  6.22 & 36 & 0.$''50\times0.''47$ (P.A. 45$^{\circ}$)&8\\
7&DCN &5$-$4 & 362.0457535 & 134 & 52 & 0.$''48\times0.46$ (P.A. 71$^{\circ}$)&5\\
\hline
\end{tabular}
}
\begin{flushleft}
\tablecomments{
$^a$ Line parameters are taken from CDMS \citep{Endres et al.(2016)}. 
The rms and the beam size are based on the observation data (2013.1.01157.S and 2015.1.01380.S).\\
$^b$ ALMA receiver band.}
\end{flushleft}
\end{table}
\end{longrotatetable}
\newpage

\begin{longrotatetable}
\begin{deluxetable*}{lllrrrrrrrrrrrrrrrrrl}
\tablecaption{Correlation Matrix for the Whole Structure \label{corr_outflow}}
\tablewidth{800pt}
\tabletypesize{\scriptsize}
\tablehead{
\colhead{Molecular spices}&\colhead {\ccchh} &\colhead{\ccchh} & \colhead{CCH}&\colhead{CCH}  &\colhead{CCH}&\colhead{CCH}&
\colhead{\meta}&\colhead{CS}&\colhead{DCN}&
\colhead{\hhco }&\colhead{\hhco }&\colhead{\hhco }&\colhead{\hhco }&\colhead{\hhco }&\colhead{SO}&\colhead{Continuum}\\
\colhead{Transition$^a$}&\colhead{ ($10-9$)}&\colhead{($9-8$)}&\colhead{($3-2$)}&\colhead{($4-3$ a)}&\colhead{($4-3$ b)}&\colhead{($4-3$ c)}&
\colhead{}&\colhead{}&\colhead{}&
\colhead{$K_a$=0}&\colhead{$K_a$=1}&\colhead{$K_a$=2}&\colhead{$K_a$=3$_u$}&\colhead{$K_a$=3$_l$}&\colhead{}&\colhead{(Cycle3)}
}
\startdata
\ccchh\ ($10-9$)	&	1	&	0.0701	&	0.2025	&	0.1514	&	0.0125	&	0.0806	&	-0.5217	&	0.0407	&	-0.0718	&	0.0229	&	-0.0012	&	-0.1951	&	-0.3934	&	-0.5902	&	-0.5435	&	-0.1789\\
\ccchh\ ($9-8$)	&	0.0701	&	1	&	0.3635	&	0.4171	&	0.3925	&	0.3666	&	-0.7422	&	0.0419	&	-0.5622	&	0.268	&	0.1289	&	0.0137	&	-0.6535	&	-0.4883	&	-0.7263	&	-0.1822\\
CCH ($3-2$)	&	0.2025	&	0.3635	&	1	&	-0.0859	&	0.6308	&	0.5969	&	0.3057	&	0.5005	&	0.4118	&	0.5671	&	0.3846	&	0.4195	&	0.3982	&	0.2723	&	0.4117	&	0.479\\
CCH ($4-3$ a)	&	0.1514	&	0.4171	&	-0.0859	&	1	&	0.3216	&	0.2884	&	-0.2207	&	0.1013	&	-0.1804	&	0.135	&	-0.058	&	0.0228	&	0.0936	&	-0.1387	&	-0.2487	&	-0.0685\\
CCH ($4-3$ b)	&	0.0125	&	0.3925	&	0.6308	&	0.3216	&	1	&	0.9726	&	0.1568	&	0.3827	&	0.548	&	0.6669	&	0.3183	&	0.335	&	0.5353	&	0.2789	&	0.0086	&	0.5312\\
CCH ($4-3$ c)	&	0.0806	&	0.3666	&	0.5969	&	0.2884	&	0.9726	&	1	&	0.0176	&	0.3561	&	0.5355	&	0.6511	&	0.3314	&	0.326	&	0.5103	&	0.2236	&	-0.0362	&	0.5254\\
\meta	&	-0.5217	&	-0.7422	&	0.3057	&	-0.2207	&	0.1568	&	0.0176	&	1	&	0.3185	&	0.1145	&	0.3633	&	0.3024	&	0.2747	&	0.3566	&	0.2606	&	0.5191	&	0.6838\\
CS	&	0.0407	&	0.0419	&	0.5005	&	0.1013	&	0.3827	&	0.3561	&	0.3185	&	1	&	0.2621	&	0.5845	&	0.5144	&	0.4508	&	0.2112	&	0.1348	&	0.0559	&	0.3402\\
DCN	&	-0.0718	&	-0.5622	&	0.4118	&	-0.1804	&	0.548	&	0.5355	&	0.1145	&	0.2621	&	1	&	0.5817	&	0.6523	&	0.4827	&	0.6319	&	0.2087	&	0.018	&	0.6525\\
\hhco ($K_a$=0)	&	0.0229	&	0.268	&	0.5671	&	0.135	&	0.6669	&	0.6511	&	0.3633	&	0.5845	&	0.5817	&	1	&	0.794	&	0.7737	&	0.6021	&	0.541	&	0.5379	&	0.6023\\
\hhco ($K_a$=1)	&	-0.0012	&	0.1289	&	0.3846	&	-0.058	&	0.3183	&	0.3314	&	0.3024	&	0.5144	&	0.6523	&	0.794	&	1	&	0.777	&	0.5855	&	0.6165	&	0.6532	&	0.4361\\
\hhco ($K_a$=2)	&	-0.1951	&	0.0137	&	0.4195	&	0.0228	&	0.335	&	0.326	&	0.2747	&	0.4508	&	0.4827	&	0.7737	&	0.777	&	1	&	0.6729	&	0.6473	&	0.4643	&	0.843\\
\hhco ($K_a$=3${_u}$)	&	-0.3934	&	-0.6535	&	0.3982	&	0.0936	&	0.5353	&	0.5103	&	0.3566	&	0.2112	&	0.6319	&	0.6021	&	0.5855	&	0.6729	&	1	&	0.6974	&	0.4244	&	0.8205\\
\hhco ($K_a$=3${_l}$)	&	-0.5902	&	-0.4883	&	0.2723	&	-0.1387	&	0.2789	&	0.2236	&	0.2606	&	0.1348	&	0.2087	&	0.541	&	0.6165	&	0.6473	&	0.6974	&	1	&	0.355	&	0.6219\\
SO	&	-0.5435	&	-0.7263	&	0.4117	&	-0.2487	&	0.0086	&	-0.0362	&	0.5191	&	0.0559	&	0.018	&	0.5379	&	0.6532	&	0.4643	&	0.4244	&	0.355	&	1	&	0.3517\\
Continuum	&	-0.1789	&	-0.1822	&	0.479	&	-0.0685	&	0.5312	&	0.5254	&	0.6838	&	0.3402	&	0.6525	&	0.6023	&	0.4361	&	0.843	&	0.8205	&	0.6219	&	0.3517	&	1\\
\hline
Rms$\times$3	&18& 18 & 30 & 12 & 30 & 60 & 30 & 30 & 90 & 30 & 90 & 30& 30& 30& 30& 1.2
\enddata
\tablecomments{
\ccchh\ ($10-9$), ($9-8$), CCH ($3-2$), ($4-3$ a), ($4-3$ b), ($4-3$ c), \hhco\ ($K_a$=0), ($K_a$=1), ($K_a$=2), ($K_a$=3${_u}$), and ($K_a$=3${_l}$) denote \ccchh\ (10$_{0,10}-9_{1,9}$ and $10_{1,10}-9_{0,9}$), \ccchh\ (9$_{1,8}-8_{2,7}$ and 9$_{2,8}-8_{1,7}$), CCH ($N=3-2, J=7/2-5/2, F=4-3$ and $3-2$), ($N=4-3, J=7/2-5/2, F=3-3$), ($N=4-3, J=7/2-5/2, F=4-3$ and $ 3-2$), ($N=4-3, J=9/2-7/2, F=5-4$ and $4-3$), \hhco\ (5$_{0,5}-4_{0,4}$), (5$_{1,5}-4_{1,4}$), (5$_{2,4}-4_{2,3}$), (5$_{3,2}-4_{3,1}$), and (5$_{3,3}-4_{3,2}$), respectively.
The units of rms are \mjybeam\kms\ and \mjybeam\ for the molecular lines and the continuum data, respectively.
}
\end{deluxetable*}
\end{longrotatetable}

\begin{longrotatetable}
\begin{deluxetable*}{lllrrrrrrrrrrrrrrrrrl}
\tablecaption{Correlation Matrix for the Disk/Envelope \label{corr_protostar}}
\tablewidth{800pt}
\tabletypesize{\scriptsize}
\tablehead{
\colhead{Molecular spices}&\colhead {\ccchh}  & \colhead{CCH}&\colhead{CCH}  &\colhead{CCH}&\colhead{CCH}&
\colhead{\meta}&\colhead{CS}&\colhead{DCN}&
\colhead{\hhco }&\colhead{\hhco }&\colhead{\hhco }&\colhead{\hhco }&\colhead{\hhco }&\colhead{SO}&\colhead{Continuum}\\
\colhead{Transition$^a$}&\colhead{ ($10-9$)}&\colhead{($3-2$)}&\colhead{($4-3$ a)}&\colhead{($4-3$ b)}&\colhead{($4-3$ c)}&
\colhead{}&\colhead{}&\colhead{}&
\colhead{$K_a$=0}&\colhead{$K_a$=1}&\colhead{$K_a$=2}&\colhead{$K_a$=3$_u$}&\colhead{$K_a$=3$_l$}&\colhead{}&\colhead{(Cycle3)}
}\startdata
\ccchh\ ($10-9$) & 1 & 0.0113 & 0.1657 & 0.5656 & 0.5949 & -0.429 & 0.4517 & -0.1288 & 0.2988 & 0.1921 & -0.3543 & -0.6046 & -0.5911 & -0.6924 & -0.198 \\
CCH ($3-2$) & 0.0113 & 1 & -0.2787 & 0.6607 & 0.7061 & 0.2147 & 0.5318 & 0.4118 & 0.6789 & 0.6262 & 0.4443 & 0.3952 & 0.3602 & 0.1209 & 0.4145 \\
CCH ($4-3$ a) & 0.1657 & -0.2787 & 1 & 0.1107 & 0.0151 & -0.2921 & -0.1203 & -0.1804 & -0.2132 & -0.2431 & -0.063 & 0.0539 & -0.2268 & -0.2487 & -0.2696 \\
CCH ($4-3$ b) & 0.5656 & 0.6607 & 0.1107 & 1 & 0.9888 & 0.3357 & 0.8041 & 0.3622 & 0.9099 & 0.6555 & 0.4827 & 0.5609 & 0.5075 & 0.0518 & 0.4411 \\
CCH ($4-3$ c) & 0.5949 & 0.7061 & 0.0151 & 0.9888 & 1 & 0.1952 & 0.8226 & 0.4028 & 0.9089 & 0.6669 & 0.466 & 0.4068 & 0.427 & -0.0568 & 0.4256 \\
\meta & -0.429 & 0.2147 & -0.2921 & 0.3357 & 0.1952 & 1 & 0.0393 & 0.1145 & 0.6638 & 0.7567 & 0.7953 & 0.4929 & 0.7601 & 0.8797 & 0.7287 \\
CS & 0.4517 & 0.5318 & -0.1203 & 0.8041 & 0.8226 & 0.0393 & 1 & 0.0297 & 0.7369 & 0.6131 & 0.2341 & 0.174 & 0.2541 & -0.0589 & 0.1175 \\
DCN & -0.1288 & 0.4118 & -0.1804 & 0.3622 & 0.4028 & 0.1145 & 0.0297 & 1 & 0.472 & 0.5036 & 0.3842 & 0.6319 & 0.2087 & 0.018 & 0.5621 \\
\hhco ($K_a$=0) & 0.2988 & 0.6789 & -0.2132 & 0.9099 & 0.9089 & 0.6638 & 0.7369 & 0.472 & 1 & 0.8341 & 0.7556 & 0.7234 & 0.6727 & 0.5264 & 0.7105 \\
\hhco ($K_a$=1) & 0.1921 & 0.6262 & -0.2431 & 0.6555 & 0.6669 & 0.7567 & 0.6131 & 0.5036 & 0.8341 & 1 & 0.8873 & 0.7993 & 0.7796 & 0.6823 & 0.8143 \\
\hhco ($K_a$=2) & -0.3543 & 0.4443 & -0.063 & 0.4827 & 0.466 & 0.7953 & 0.2341 & 0.3842 & 0.7556 & 0.8873 & 1 & 0.7817 & 0.8834 & 0.8691 & 0.9289 \\
\hhco ($K_a$=3$_u$) & -0.6046 & 0.3952 & 0.0539 & 0.5609 & 0.4068 & 0.4929 & 0.174 & 0.6319 & 0.7234 & 0.7993 & 0.7817 & 1 & 0.7436 & 0.5551 & 0.8988 \\
\hhco ($K_a$=3$_l$) & -0.5911 & 0.3602 & -0.2268 & 0.5075 & 0.427 & 0.7601 & 0.2541 & 0.2087 & 0.6727 & 0.7796 & 0.8834 & 0.7436 & 1 & 0.8139 & 0.8934 \\
SO & -0.6924 & 0.1209 & -0.2487 & 0.0518 & -0.0568 & 0.8797 & -0.0589 & 0.018 & 0.5264 & 0.6823 & 0.8691 & 0.5551 & 0.8139 & 1 & 0.8104 \\
Continuum & -0.198 & 0.4145 & -0.2696 & 0.4411 & 0.4256 & 0.7287 & 0.1175 & 0.5621 & 0.7105 & 0.8143 & 0.9289 & 0.8988 & 0.8934 & 0.8104 & 1 \\
\hline
\enddata
\tablecomments{
\ccchh\ ($10-9$), CCH ($3-2$), ($4-3$ a), ($4-3$ b), ($4-3$ c), \hhco\ ($K_a$=0), ($K_a$=1), ($K_a$=2), ($K_a$=3${_u}$), and ($K_a$=3${_l}$) denote \ccchh\ (10$_{0,10}-9_{1,9}$ and $10_{1,10}-9_{0,9}$), CCH ($N=3-2, J=7/2-5/2, F=4-3$ and $3-2$), ($N=4-3, J=7/2-5/2, F=3-3$), ($N=4-3, J=7/2-5/2, F=4-3$ and $ 3-2$), ($N=4-3, J=9/2-7/2, F=5-4$ and $4-3$), \hhco\ (5$_{0,5}-4_{0,4}$), (5$_{1,5}-4_{1,4}$), (5$_{2,4}-4_{2,3}$), (5$_{3,2}-4_{3,1}$), and (5$_{3,3}-4_{3,2}$), respectively.
The units of rms are \mjybeam\kms\ and \mjybeam\ for the molecular lines and the continuum data, respectively.}
\end{deluxetable*}
\end{longrotatetable}

\begin{longrotatetable}
\begin{deluxetable*}{lllrrrrrrrrrrrrrrrrl}
\tablecaption{Eigen Vectors of the Principal Components and their Eigen Values in the Analysis of the Whole Structure \label{vec_outflow}}
\tablewidth{800pt}
\tabletypesize{\scriptsize}
\tablehead{
\colhead{Principal Component}&\colhead {PC1} &\colhead{PC2} & \colhead{PC3}&\colhead{PC4}  &
\colhead{PC5}&\colhead{PC6}&\colhead{PC7}&\colhead{PC8}&
\colhead{PC9}&\colhead{PC10}&\colhead{PC11}&\colhead{PC12}&\colhead{PC13}&\colhead{PC14}&\colhead{PC15}&\colhead{PC16}
}
\startdata
\ccchh\ (10-9)&	-0.098	&	0.298	&	-0.131	&	-0.559	&	0.215	&	0.234	&	0.188	&	0.46	&	-0.165	&	0.31	&	-0.025	&	0.077	&	0.018	&	0.102	&	0.018	&	0.288\\
\ccchh\ (9-8)	&	-0.077	&	0.514	&	0.385	&	0.213	&	-0.014	&	-0.269	&	-0.188	&	0.024	&	0.234	&	-0.043	&	0.125	&	0.062	&	0.178	&	-0.093	&	-0.111	&	0.546\\
CCH ($3-2$)	&	0.249	&	0.203	&	0.136	&	-0.267	&	-0.316	&	-0.318	&	0.178	&	0.263	&	-0.296	&	-0.404	&	0.363	&	-0.188	&	-0.041	&	-0.161	&	0.099	&	-0.224\\
CCH ($4-3$ a)&	-0.006	&	0.288	&	-0.02	&	0.436	&	0.001	&	0.666	&	0.293	&	0.074	&	0.038	&	-0.154	&	0.287	&	-0.138	&	-0.242	&	-0.069	&	-0.04	&	-0.022\\
CCH ($4-3$ b)&	0.257	&	0.328	&	-0.221	&	0.15	&	-0.242	&	-0.128	&	0.121	&	-0.145	&	-0.024	&	0.177	&	0.012	&	-0.052	&	0.141	&	0.762	&	-0.021	&	-0.089\\
CCH ($4-3$ c)&	0.244	&	0.345	&	-0.249	&	0.112	&	-0.151	&	-0.155	&	0.139	&	-0.122	&	-0.034	&	0.16	&	-0.345	&	0.433	&	-0.367	&	-0.376	&	0.236	&	-0.024\\
\meta&	0.201	&	-0.299	&	0.033	&	-0.086	&	-0.58	&	0.227	&	-0.127	&	0.047	&	0.228	&	0.337	&	0.302	&	0.044	&	0.057	&	-0.073	&	0.345	&	0.264\\
CS&	0.203	&	0.147	&	0.224	&	-0.277	&	-0.192	&	0.423	&	-0.405	&	-0.362	&	-0.385	&	-0.217	&	-0.235	&	0.073	&	-0.02	&	0.05	&	-0.178	&	0.093\\
DCN&	0.268	&	0.013	&	-0.461	&	-0.279	&	0.253	&	-0.076	&	-0.054	&	-0.359	&	0.286	&	-0.206	&	0.22	&	-0.311	&	-0.232	&	-0.054	&	-0.033	&	0.331\\
\hhco ($K_a$=0)&	0.339	&	0.155	&	0.224	&	-0.041	&	0.078	&	0.063	&	0.122	&	-0.079	&	0.138	&	0.391	&	-0.28	&	-0.568	&	0.262	&	-0.302	&	-0.051	&	-0.208\\
\hhco ($K_a$=1)&	0.304	&	0.014	&	0.318	&	-0.155	&	0.386	&	0.041	&	0.091	&	-0.244	&	0.117	&	0.178	&	0.455	&	0.476	&	0.053	&	0.026	&	-0.048	&	-0.279\\
\hhco ($K_a$=2)	&	0.32	&	-0.007	&	0.216	&	0.061	&	0.297	&	0.079	&	-0.252	&	0.328	&	0.217	&	-0.27	&	-0.248	&	-0.036	&	-0.141	&	0.264	&	0.556	&	-0.029\\
\hhco ($K_a$=3${_u}$)&	0.328	&	-0.124	&	-0.3	&	0.191	&	0.123	&	0.103	&	0.154	&	0.077	&	-0.182	&	-0.261	&	-0.051	&	0.208	&	0.689	&	-0.168	&	0.05	&	0.206\\
\hhco ($K_a$=3${_l}$)	&	0.264	&	-0.196	&	0.047	&	0.336	&	0.237	&	-0.162	&	-0.167	&	0.096	&	-0.591	&	0.331	&	0.182	&	-0.158	&	-0.268	&	0.005	&	-0.04	&	0.258\\
SO	&	0.211	&	-0.325	&	0.348	&	-0.053	&	-0.096	&	-0.049	&	0.601	&	-0.017	&	0.09	&	-0.141	&	-0.268	&	0.061	&	-0.214	&	0.159	&	-0.246	&	0.344\\
Continuum&	0.334	&	-0.046	&	-0.172	&	0.042	&	-0.108	&	0.015	&	-0.309	&	0.476	&	0.274	&	0.003	&	-0.035	&	0.136	&	-0.11	&	-0.04	&	-0.625	&	-0.149\\
\hline
Eigen values& 6.731& 3.243& 1.397& 1.310&  1.108&  8.830&
  7.450& 5.940&  4.710&  2.960&   1.660& 8.100&
  1.300&  4.000& -8.900& -9.540\\
Contribution ratio (\%) &42.1& 20.3 & 8.7 & 8.2 & 6.9 & 5.5&  4.7 & 3.7 & 2.9  &1.9 & 1.0  & 0.5&  0.1&  0.0&
 -0.6 &-6.0\\
\enddata
\begin{flushleft}
\tablecomments{
\ccchh\ ($10-9$), ($9-8$), CCH ($3-2$), ($4-3$ a), ($4-3$ b), ($4-3$ c), \hhco\ ($K_a$=0), ($K_a$=1), ($K_a$=2), ($K_a$=3${_u}$), and ($K_a$=3${_l}$) denote \ccchh\ (10$_{0,10}-9_{1,9}$ and $10_{1,10}-9_{0,9}$), \ccchh\ (9$_{1,8}-8_{2,7}$ and 9$_{2,8}-8_{1,7}$), CCH ($N=3-2, J=7/2-5/2, F=4-3$ and $3-2$), ($N=4-3, J=7/2-5/2, F=3-3$), ($N=4-3, J=7/2-5/2, F=4-3$ and $ 3-2$), ($N=4-3, J=9/2-7/2, F=5-4$ and $4-3$), \hhco\ (5$_{0,5}-4_{0,4}$), (5$_{1,5}-4_{1,4}$), (5$_{2,4}-4_{2,3}$), (5$_{3,2}-4_{3,1}$), and (5$_{3,3}-4_{3,2}$), respectively.}
\end{flushleft}
\end{deluxetable*}
\end{longrotatetable}

\begin{longrotatetable}
\begin{deluxetable*}{lllrrrrrrrrrrrrrrrrl}
\tablecaption{Eigen Vectors of the Principal Components and their Eigen Values in the Analysis of the Disk/Envelope \label{vec_protostar}}
\tablewidth{800pt}
\tabletypesize{\scriptsize}
\tablehead{
\colhead{Principal Component}&\colhead {PC1} &\colhead{PC2} & \colhead{PC3}&\colhead{PC4}  &
\colhead{PC5}&\colhead{PC6}&\colhead{PC7}&\colhead{PC8}&
\colhead{PC9}&\colhead{PC10}&\colhead{PC11}&\colhead{PC12}&\colhead{PC13}&\colhead{PC14}&\colhead{PC15}
}
\startdata
\ccchh\ ($10-9$) & -0.045 & -0.496 & -0.18 & 0.067 & 0.548 & -0.084 & -0.077 & 0.186 & 0.083 & -0.133 & 0.058 & 0.012 & -0.101 & 0.144 & -0.555 \\
CCH ($3-2$) & 0.227 & -0.189 & 0.101 & -0.331 & -0.459 & -0.693 & -0.088 & 0.092 & 0.101 & -0.166 & 0.078 & -0.043 & 0.083 & 0.1 & -0.153 \\
CCH ($4-3$ a) & -0.078 & -0.084 & 0.264 & 0.832 & -0.079 & -0.26 & -0.128 & -0.079 & -0.061 & 0.048 & 0.148 & -0.201 & 0.219 & 0.109 & 0.029 \\
CCH ($4-3$ b) & 0.263 & -0.341 & 0.013 & 0.182 & -0.057 & 0.037 & 0.38 & -0.071 & 0.014 & -0.252 & 0.398 & 0.484 & -0.17 & -0.3 & 0.23 \\
CCH ($4-3$ c) & 0.248 & -0.379 & 0.024 & 0.049 & -0.036 & -0.016 & 0.229 & 0.192 & -0.158 & 0.273 & -0.438 & -0.415 & -0.058 & -0.481 & 0.059 \\
\meta & 0.263 & 0.231 & -0.313 & 0.04 & 0.207 & -0.231 & 0.338 & -0.52 & -0.269 & -0.163 & -0.196 & 0.007 & 0.351 & -0.009 & -0.191 \\
CS & 0.177 & -0.366 & -0.259 & -0.027 & -0.35 & 0.431 & -0.377 & -0.182 & -0.044 & 0.163 & 0.072 & 0.113 & 0.463 & -0.006 & -0.147 \\
DCN & 0.178 & -0.036 & 0.664 & -0.3 & 0.283 & 0.063 & -0.01 & -0.234 & -0.259 & 0.295 & 0.331 & -0.054 & 0.1 & -0.031 & -0.143 \\
\hhco ($K_a$=0) & 0.335 & -0.174 & -0.09 & -0.004 & 0.083 & 0.039 & 0.189 & -0.276 & 0.481 & 0.292 & 0.005 & -0.213 & -0.156 & 0.517 & 0.275 \\
\hhco ($K_a$=1) & 0.341 & -0.048 & -0.085 & -0.025 & 0.213 & -0.014 & -0.551 & -0.083 & -0.34 & -0.384 & -0.005 & -0.196 & -0.248 & 0.04 & 0.392 \\
\hhco ($K_a$=2) & 0.325 & 0.152 & -0.019 & 0.165 & 0.059 & -0.205 & -0.214 & 0.171 & -0.15 & 0.478 & -0.274 & 0.593 & -0.176 & 0.107 & -0.053 \\
\hhco ($K_a$=3$_u$) & 0.297 & 0.131 & 0.408 & 0.152 & -0.163 & 0.291 & -0.068 & -0.155 & 0.311 & -0.381 & -0.374 & 0.04 & -0.193 & -0.041 & -0.378 \\
\hhco ($K_a$=3$_l$) & 0.308 & 0.191 & -0.112 & 0.091 & -0.244 & 0.25 & 0.288 & 0.398 & -0.397 & -0.037 & 0.269 & -0.219 & -0.136 & 0.358 & -0.242 \\
SO & 0.241 & 0.36 & -0.273 & 0.082 & 0.047 & -0.085 & -0.211 & -0.039 & 0.331 & 0.204 & 0.42 & -0.222 & -0.156 & -0.474 & -0.222 \\
Continuum & 0.32 & 0.153 & 0.109 & -0.016 & 0.299 & 0.049 & 0.018 & 0.504 & 0.282 & -0.162 & -0.012 & 0.04 & 0.598 & -0.018 & 0.222 \\
\hline
Eigen values& 7.918  &3.581 & 1.281 & 1.166 & 0.708 & 0.376 & 0.289 & 0.247  &0.121 & 0.109&
 -0.023& -0.033 &-0.062 &-0.13  &-0.548\\
Contribution ratio (\%) &52.8& 23.9 & 8.5 & 7.8 & 4.7&  2.5 & 1.9 & 1.6  &0.8  &0.7 &-0.2& -0.2& -0.4& -0.9&-3.7\\
\enddata
\begin{flushleft}
\tablecomments{\ccchh\ ($10-9$), CCH ($3-2$), ($4-3$ a), ($4-3$ b), ($4-3$ c), \hhco\ ($K_a$=0), ($K_a$=1), ($K_a$=2), ($K_a$=3${_u}$), and ($K_a$=3${_l}$) denote \ccchh\ (10$_{0,10}-9_{1,9}$ and $10_{1,10}-9_{0,9}$), CCH ($N=3-2, J=7/2-5/2, F=4-3$ and $3-2$), ($N=4-3, J=7/2-5/2, F=3-3$), ($N=4-3, J=7/2-5/2, F=4-3$ and $3-2$), ($N=4-3, J=9/2-7/2, F=5-4$ and $4-3$), \hhco\ (5$_{0,5}-4_{0,4}$), (5$_{1,5}-4_{1,4}$), (5$_{2,4}-4_{2,3}$), (5$_{3,2}-4_{3,1}$), and (5$_{3,3}-4_{3,2}$), respectively.
}
\end{flushleft}
\end{deluxetable*}
\end{longrotatetable}

\begin{table}[h!]
\centering
\caption{Positions of the \hhco\ Blobs \label{position}}
\scalebox{1.0}{
\begin{tabular}{ccc}
\hline\hline
Position & R.A. (J2000) & Decl. (J2000)\\
\hline
A & 15\fh43\fm02\fs47   & $-$34\arcdeg09\arcmin04\farcs94 \\
B  & 15\fh43\fm02\fs25 & $-$34\arcdeg09\arcmin05\farcs55\\
C & 15\fh43\fm02\fs07  & $-$34\arcdeg09\arcmin09\farcs23\\
D &  15\fh43\fm02\fs08  & $-$34\arcdeg09\arcmin10\farcs63\\
Center & 15\fh43\fm02\fs24 & $-$34\arcdeg 09\arcmin 06\farcs83\\
\hline
\end{tabular}
}
\begin{flushleft}
\end{flushleft}
\end{table}

\begin{table}[h!]
\centering
\caption{Line Parameters at the \hhco\ Blobs  \label{lines}}
\scalebox{1.0}{
\begin{tabular}{ccccc}
\hline\hline
Position  & Transition & $T_{\rm peak}$ (K) & $V_{\rm LSRK} $(\kms) & FWHM (\kms) \\
\hline
A&$K_a$=0  & 6.5 (0.15) & 4.6 (0.01) & 0.9 (0.04) \\
  &$K_a$=1   & 10 (0.22) & 4.6 (0.02)  & 1.1 (0.06) \\ 
  &$K_a$=2   & 2.0 (0.08) & 4.6 (0.03) & 0.9 (0.04)\\
  &$K_a$=3${_u}$   & 1.3 (0.09) & 4.6 (0.05) & 0.9 (0.07)\\
  &$K_a$=3${_l}$&1.5 (0.07)& 4.5 (0.03)& 0.9 (0.04)\\
B&$K_a$=0   & 3.4 (0.21) & 3.2 (0.08) & 1.4 (0.06)\\
  &$K_a$=1  & 5.6 (0.44) & 3.5 (0.30) & 1.5(0.06)\\
  &$K_a$=2  & 0.6 (0.08) & 3.2 (0.21) & 1.5 (0.14)\\
  &$K_a$=3${_u}$  & 0.6 (0.08) & 2.8 (0.07) & 1.1 (0.14)\\
   &$K_a$=3${_l}$&0.3 (0.05)& 3.4 (0.13)& 1.1 (0.22)\\
C&$K_a$=0  & 1.7 (0.15) & 1.0 (0.24) & 1.9 (0.14)\\
  &$K_a$=1  & 4.5 (0.33) & 1.0 (0.24) & 2.1 (0.11)\\
  &$K_a$=2  & 0.6 (0.04) & 0.7 (0.16) & 1.6 (0.14)\\
  &$K_a$=3${_u}$  & 0.3 (0.09) & 0.6 (0.11) & 0.9 (0.31)\\
  &$K_a$=3${_l}$&0.4 (0.05)& 0.9 (0.12) & 1.4 (0.20)\\
D&$K_a$=0  & 4.5 (0.22) & 2.1 (0.04) & 1.1 (0.04)\\
  &$K_a$=1  & 7.6 (0.39) & 2.0 (0.04) & 1.2 (0.05)\\
  &$K_a$=2  & 1.1 (0.05) & 2.1 (0.04) & 1.0 (0.05)\\
  &$K_a$=3${_u}$  & 0.5 (0.05) & 2.0 (0.10) & 1.3 (0.16)\\
  &$K_a$=3${_l}$&0.6 (0.06)&2.1 (0.06)& 0.9 (0.10)\\
Center&$K_a=$0  & 6.0 (0.13)   &  3.0 (0.01) & 0.9 (0.02)\\
  &$K_a$=1  &          8.0  (0.46)  &  3.1 (0.04) & 1.1 (0.07)\\
  &$K_a$=2  &	     1.9  (0.08)  &  3.0 (0.03) & 0.9 (0.04)\\
  &$K_a$=3${_u}$  &	     0.8  (0.06)  &  2.7 (0.09) & 1.6 (0.16)\\
  &$K_a$=3${_l}$& 0.7 (0.06)& 3.0 (0.08)& 1.4 (0.14)\\
\hline
\end{tabular}}
\begin{flushleft}
\tablecomments{Measured for a circular area in Figure \ref{position} with a diameter of 155 au. 'Center' denotes the protostar position. The  line parameters are obtained by using gaussian-fitting. The numbers in parentheses represent the gaussian-fitting errors. $K_a$=0, $K_a$=1, $K_a$=2, $K_a$=3${_u}$, and $K_a$=3${_l}$ denote \hhco\ (5$_{0,5}-4_{0,4}$), (5$_{1,5}-4_{1,4}$), (5$_{2,4}-4_{2,3}$), (5$_{3,2}-4_{3,1}$), and (5$_{3,3}-4_{3,2}$), respectively.}
\end{flushleft}
\end{table}

\begin{table}[h!]
\centering
\caption{Gas Kinematic Temperatures and Column Densities of the Blobs\label{tempvalue}}
\scalebox{1.0}{
\begin{tabular}{ccccccc}
\hline \hline
Blob& Column density (10$^{14}$ cm$^{-2}$) & T$_{gas}$ (K) & ortho/para &\\
\hline
Center &  0.87$\pm$ 0.05 &55 $\pm$ 2& 1.7$\pm$ 0.13&\\
A & 0.83$\pm$ 0.0 &63 $\pm$ 2& 2.0 $\pm$ 0.09& \\  
B &0.66$\pm$0.06 &43$\pm$3& 2.3$\pm$ 0.26& \\
C & 0.48 $\pm$0.04 & 54$\pm$ 4& 3.3$\pm$ 0.39&\\ 
D & 0.78$\pm$0.06 & 45$\pm$ 2& 2.4$\pm$0.2 &\\
\hline
\end{tabular}
}
\begin{flushleft}
\tablecomments{
The H$_{2}$ density is assumed to be 10$^6$ cm$^{-3}$. The derived values are not much different even if the H$_{2}$ densitiy is 10$^5$ cm$^{-3}$ or 10$^7$ cm$^{-3}$.
The errors are derived from the least-squares analysis on the intensities of five \hhco\ lines.
'Center' denotes the protostar position.}
\end{flushleft}
\end{table}

\newpage

\begin{figure}
\centering
\includegraphics[scale=0.5]{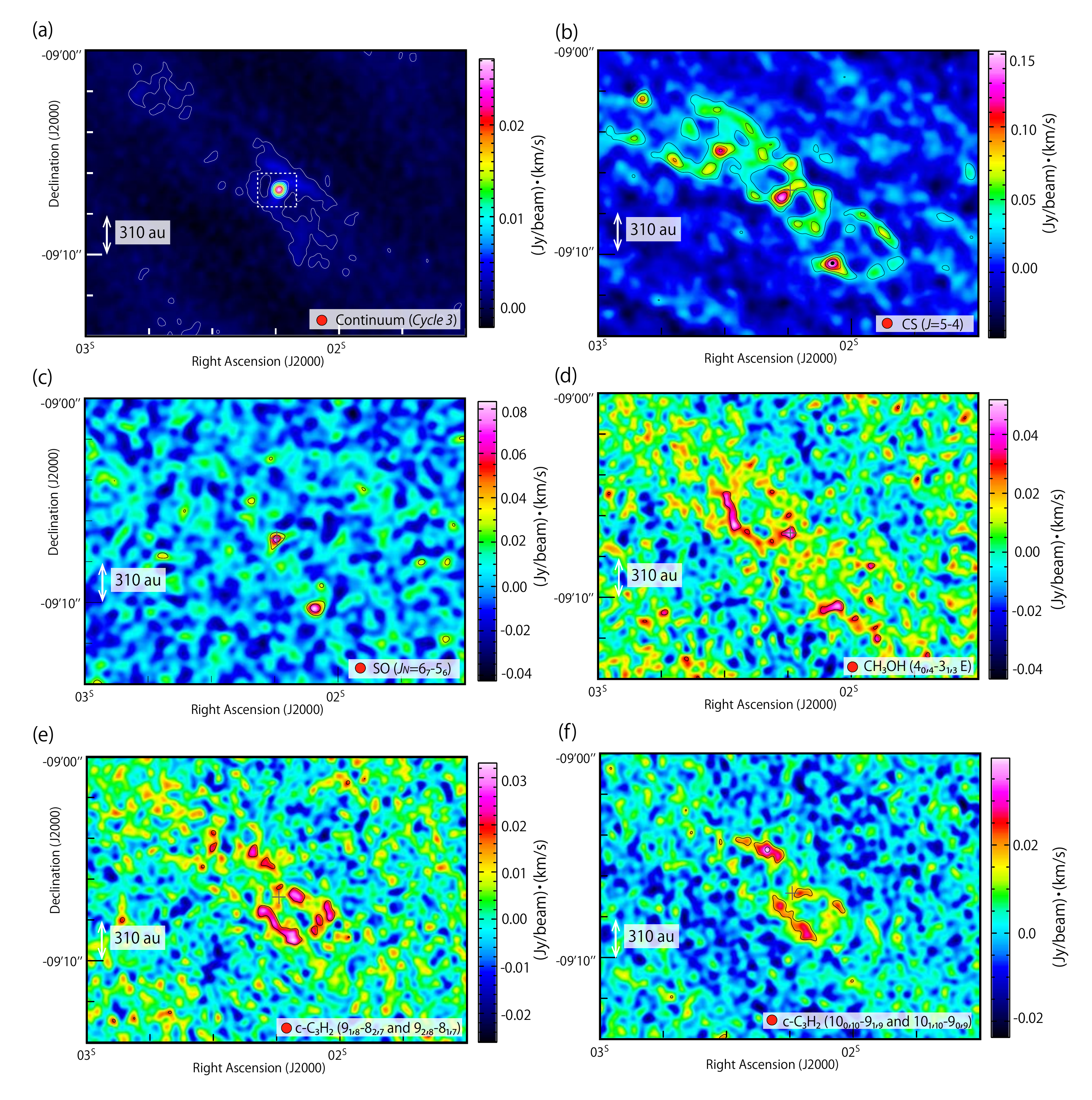}
\caption{(a) The 0.8 mm continuum image and (b-p) the moment 0 maps of 15 molecular lines. The area enclosed by a white dashed line shows the blow-up area for Figure \ref{protostar}. The red circle in the bottom right corner of each map shows the beam size, which is unified to be 0.5$''\times$0.5$''$. The cross marks show the continuum peak position.  Contour levels are every $\sigma$, every 2$\sigma$, every 3$\sigma$, every 5$\sigma$, and every 10$\sigma$ from 3$\sigma$ for (l, p), (n), (b-f, h-k, m), (g) and (a, o) respectively. 3$\sigma$ is listed in Table \ref{corr_outflow}.\label{outflow}}
\end{figure}
\addtocounter{figure}{-1}
\begin{figure}
\centering
\includegraphics[scale=0.5]{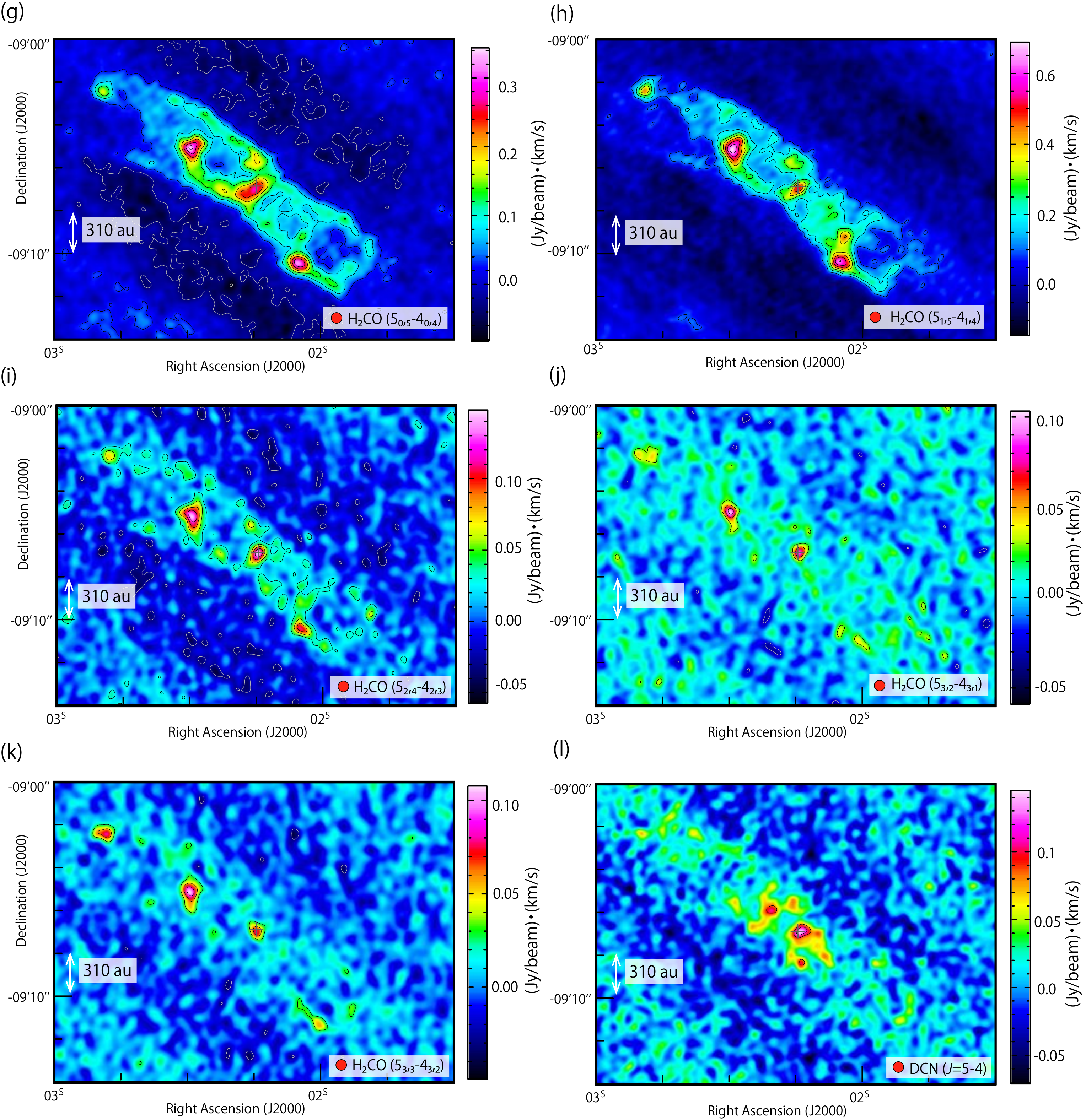}
\caption{{\bf continued.}\label{outflow}}

\end{figure}
\addtocounter{figure}{-1}
\begin{figure}
\centering
\includegraphics[scale=0.5]{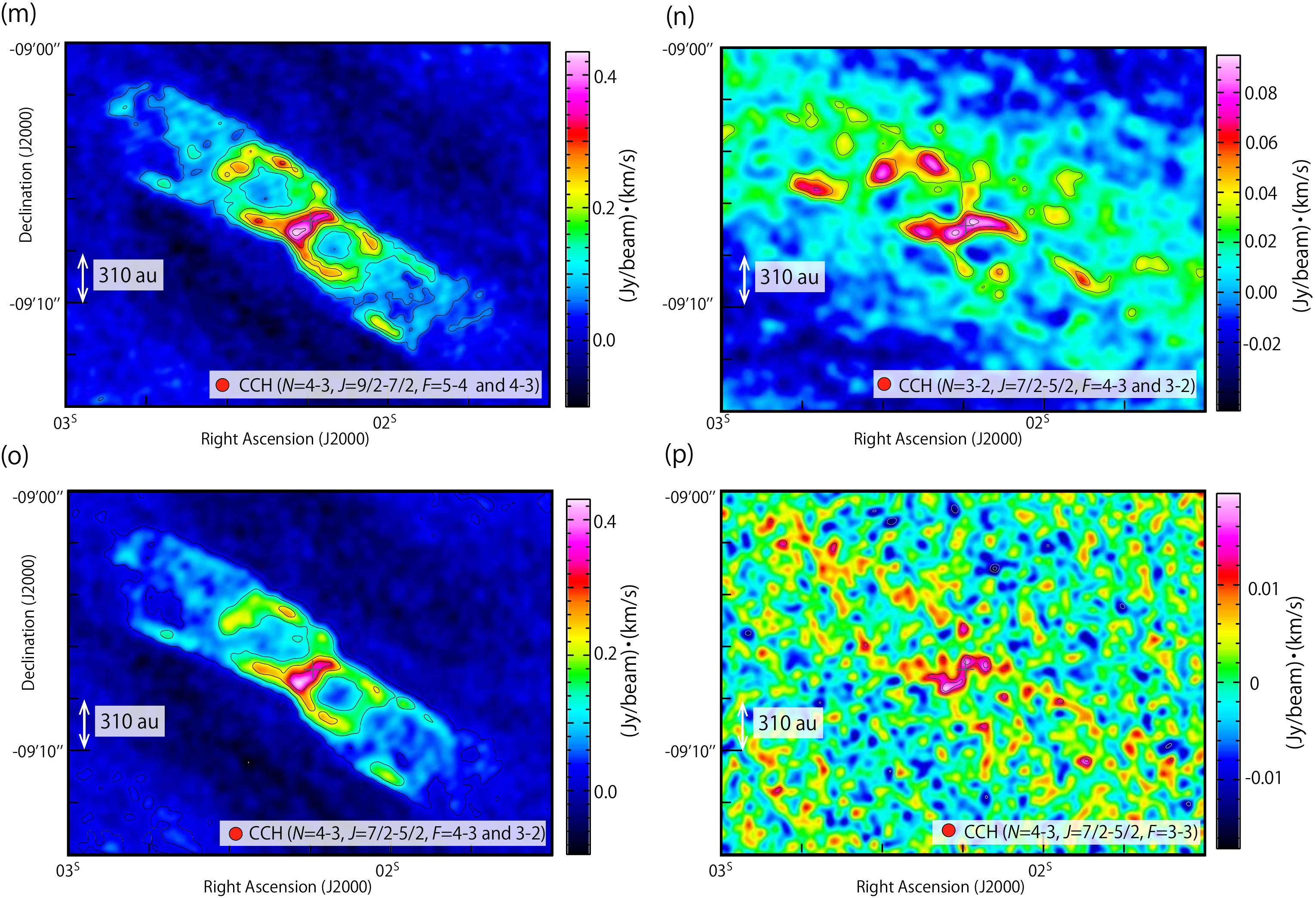}
\caption{{\bf continued.}\label{outflow}}

\end{figure}

\begin{figure}
\centering
\includegraphics[scale=1]{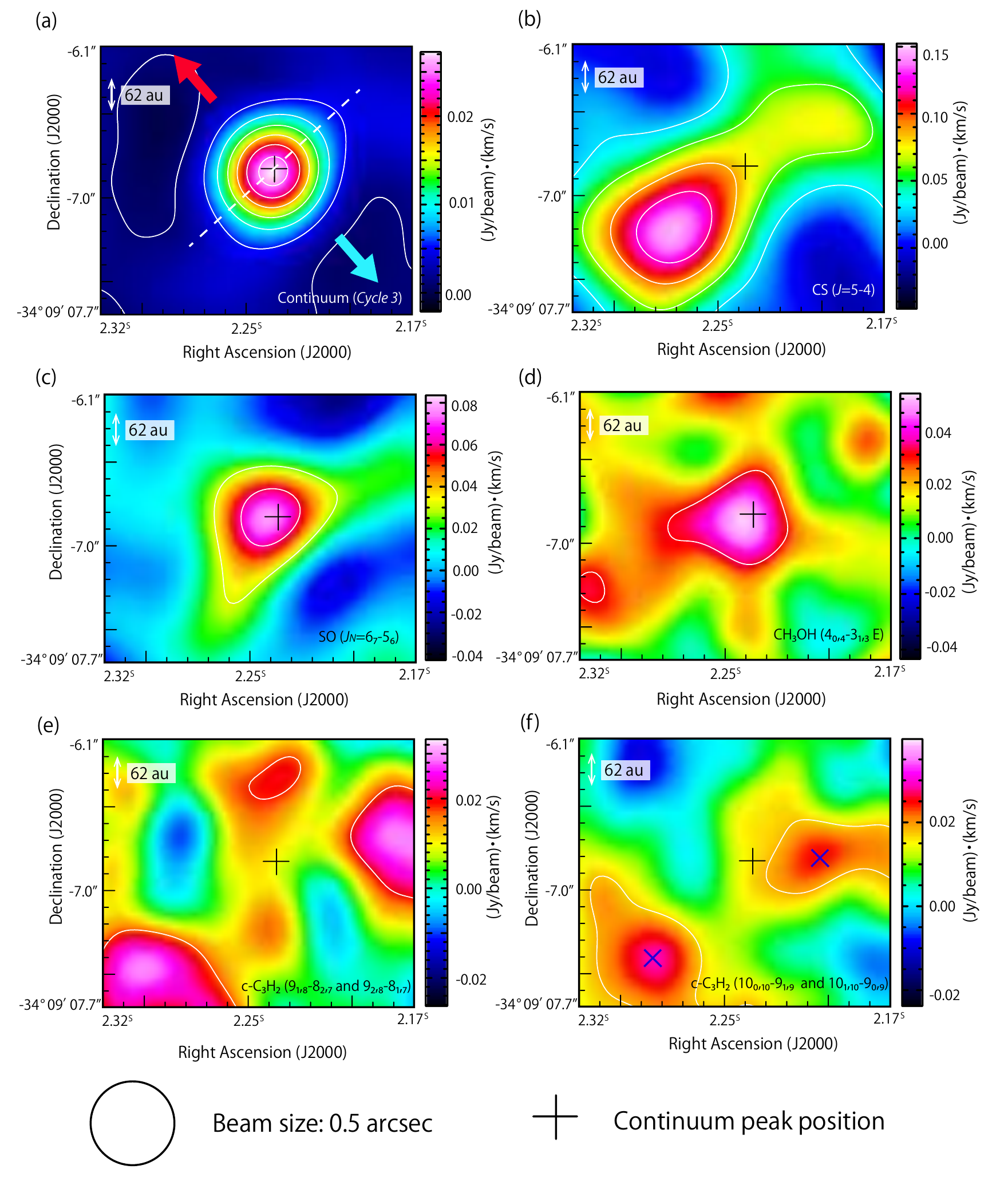}
\caption{The blow-ups of Figure 1 around the protostar. The outflow and envelope direction (P.A. 130 $\deg$) are drawn with the arrows and the dashed line, respectively, in the panel (a).  The blue cross marks in the panel (f) represent the peak positions for the calculation of the column density ratios relative to \hhco\ in Section 5. The circle in the bottom shows the beam size. The black cross marks show the continuum peak position. Contour levels are the same as those of Figure \ref{outflow}.\label{protostar}}
\end{figure}
\addtocounter{figure}{-1}
\begin{figure}
\centering
\includegraphics[scale=1]{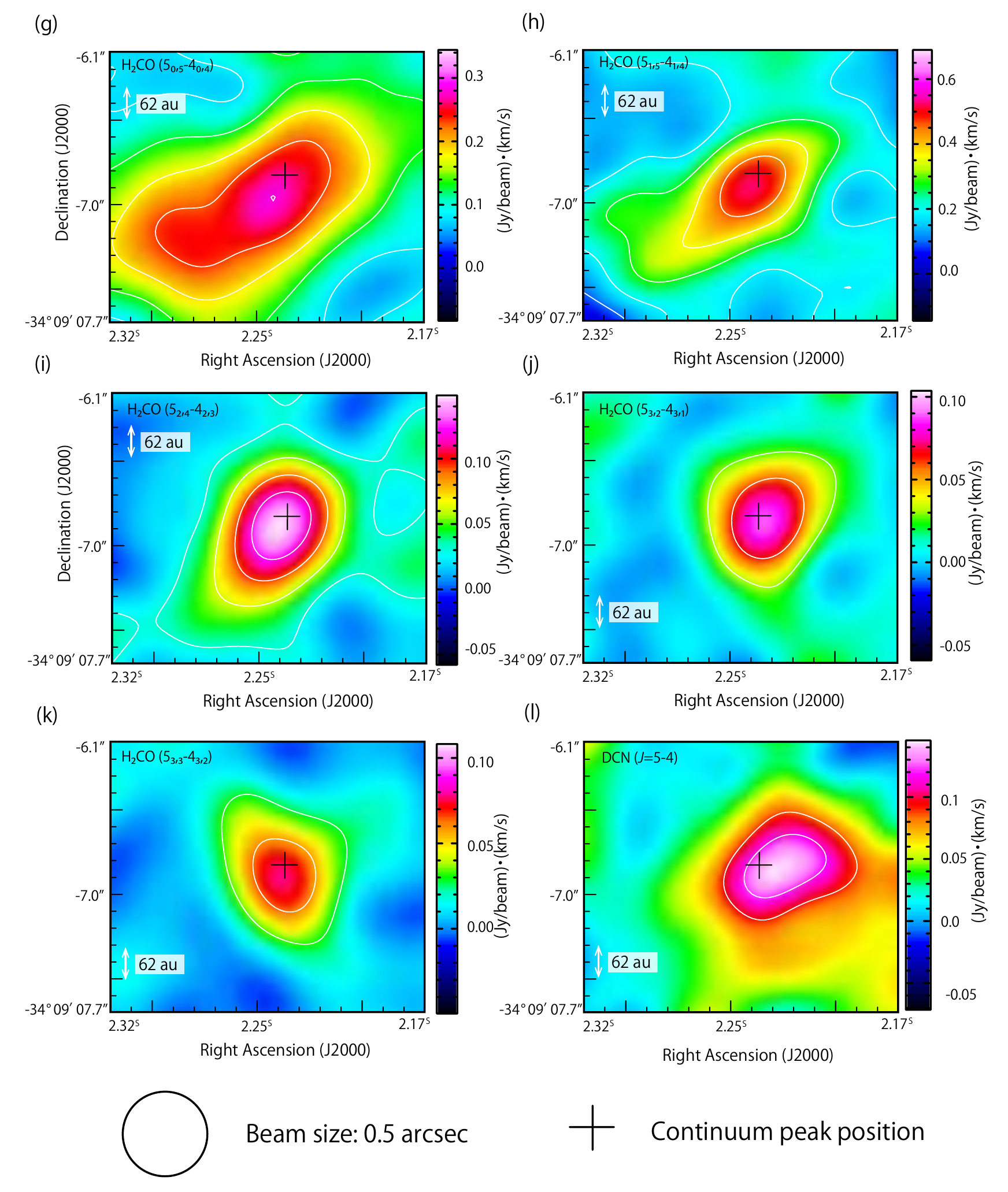}
\caption{{\bf continued.}\label{protostar}}

\end{figure}
\addtocounter{figure}{-1}
\begin{figure}
\centering
\includegraphics[scale=1]{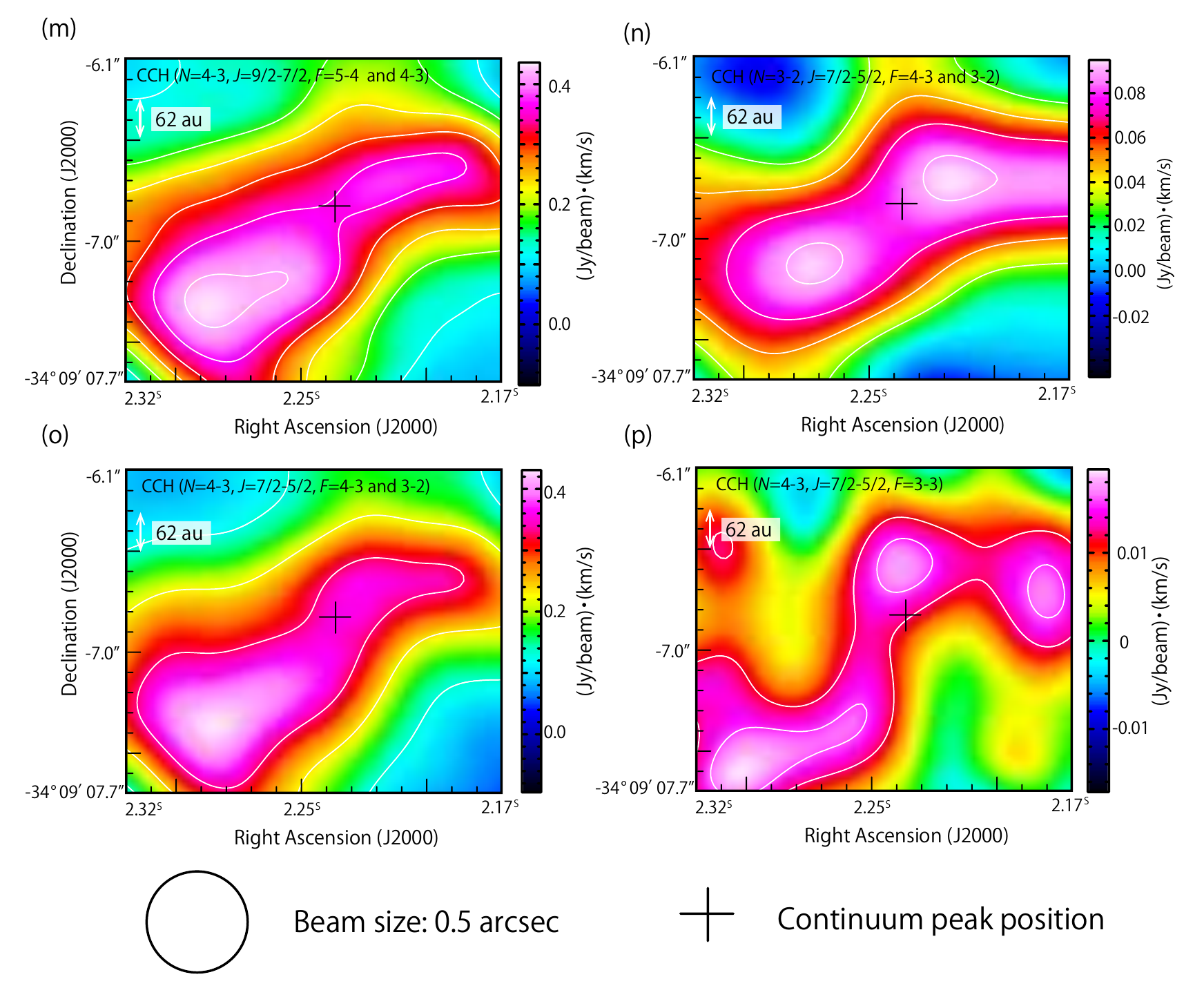}
\caption{{\bf continued.}\label{protostar}}

\end{figure}

\begin{figure}[h!]
\centering
\includegraphics[scale=0.55]{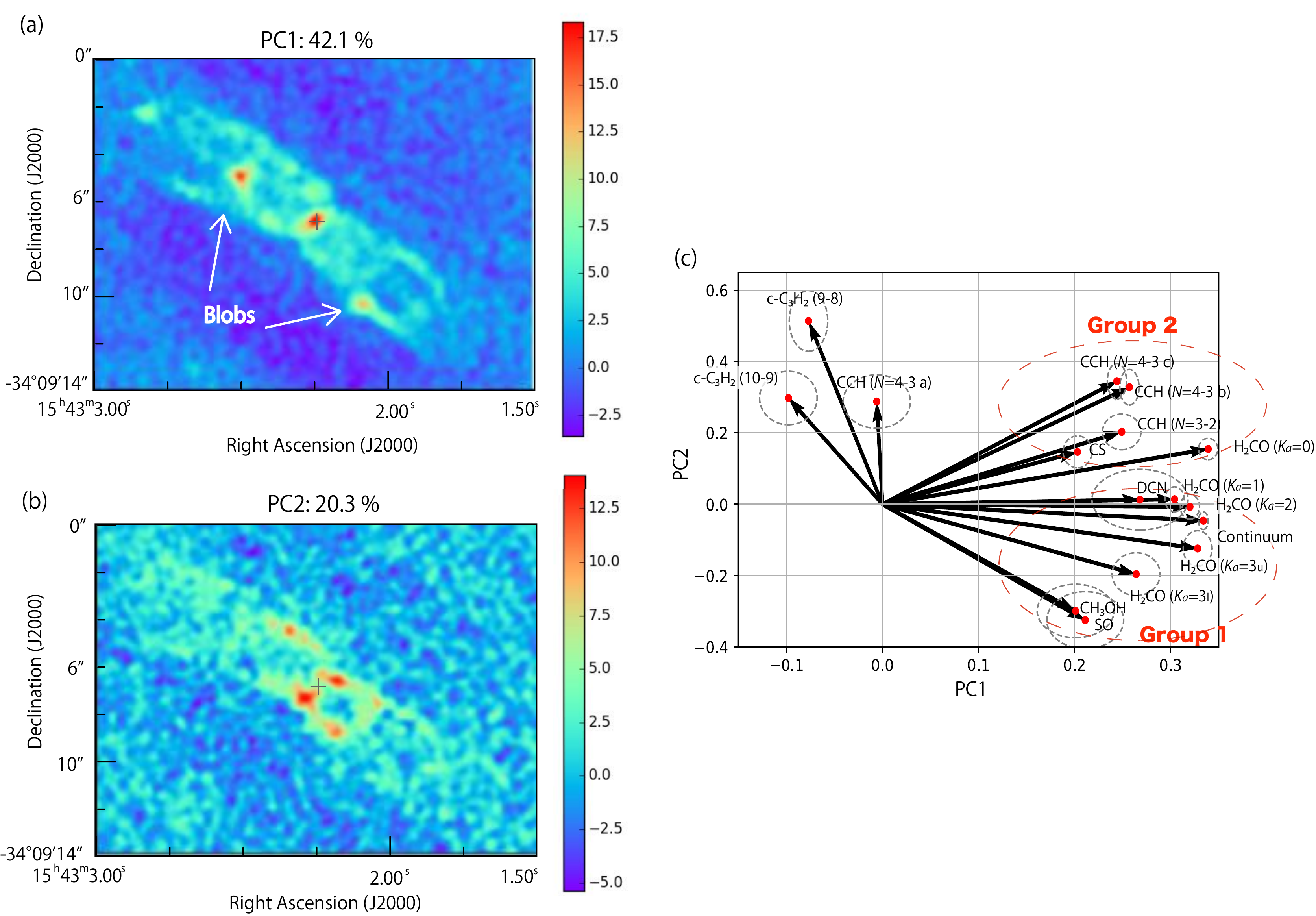}
\caption{(a,b) The principal components PC1 and PC2 for the whole-scale structure (Figure \ref{outflow}). (c) The plot of the principal components for each distribution on the PC1-PC2 plane.
The cross marks show the continuum peak position. The red dashed circles show the groups of the compact distribution (Group 1) and the extended distribution (Group 2).
The grey dashed ellipses represent the uncertainties (See Section 4.3).
Blobs in (a) is consistent with the blobs A and D in Figure \ref{temperature_region}.
\ccchh\ ($10-9$), ($9-8$), CCH ($N=3-2$), ($N=4-3$ a), ($N=4-3$ b), ($N=4-3$ c), \hhco\ ($K_a$=0), ($K_a$=1), ($K_a$=2), ($K_a$=3${_u}$), and ($K_a$=3${_l}$) denote \ccchh\ (10$_{0,10}-9_{1,9}$ and $10_{1,10}-9_{0,9}$), \ccchh\ (9$_{1,8}-8_{2,7}$ and 9$_{2,8}-8_{1,7}$), CCH ($N=3-2, J=7/2-5/2, F=4-3$ and $3-2$), ($N=4-3, J=7/2-5/2, F=3-3$), ($N=4-3, J=7/2-5/2, F=4-3$ and $ 3-2$), ($N=4-3, J=9/2-7/2, F=5-4$ and $4-3$), \hhco\ (5$_{0,5}-4_{0,4}$), (5$_{1,5}-4_{1,4}$), (5$_{2,4}-4_{2,3}$), (5$_{3,2}-4_{3,1}$), and (5$_{3,3}-4_{3,2}$), respectively.\label{pc_outflow}}
\end{figure}

\begin{figure}[h!]
\centering
\includegraphics[scale=1.0]{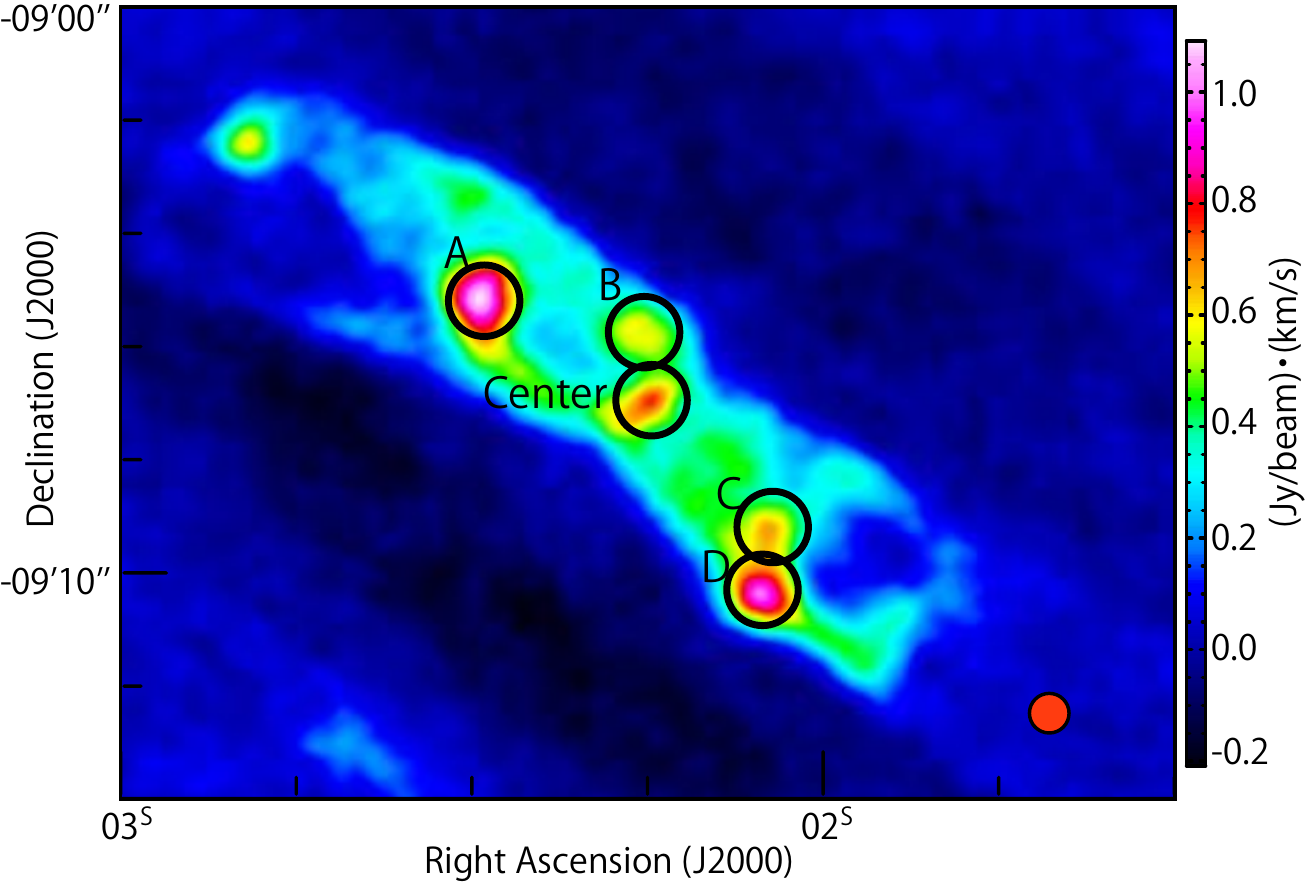}
\caption{Blobs in the outflow indicated on the moment 0 map of \hhco\ ($K_a=$1). The positions of the blobs are listed in Table \ref{position}. 
The red circle shows the beam size of 0\farcs5$\times$0\farcs5.}\label{temperature_region}
\end{figure}

\begin{figure}[h!]
\centering
\includegraphics[scale=0.55]{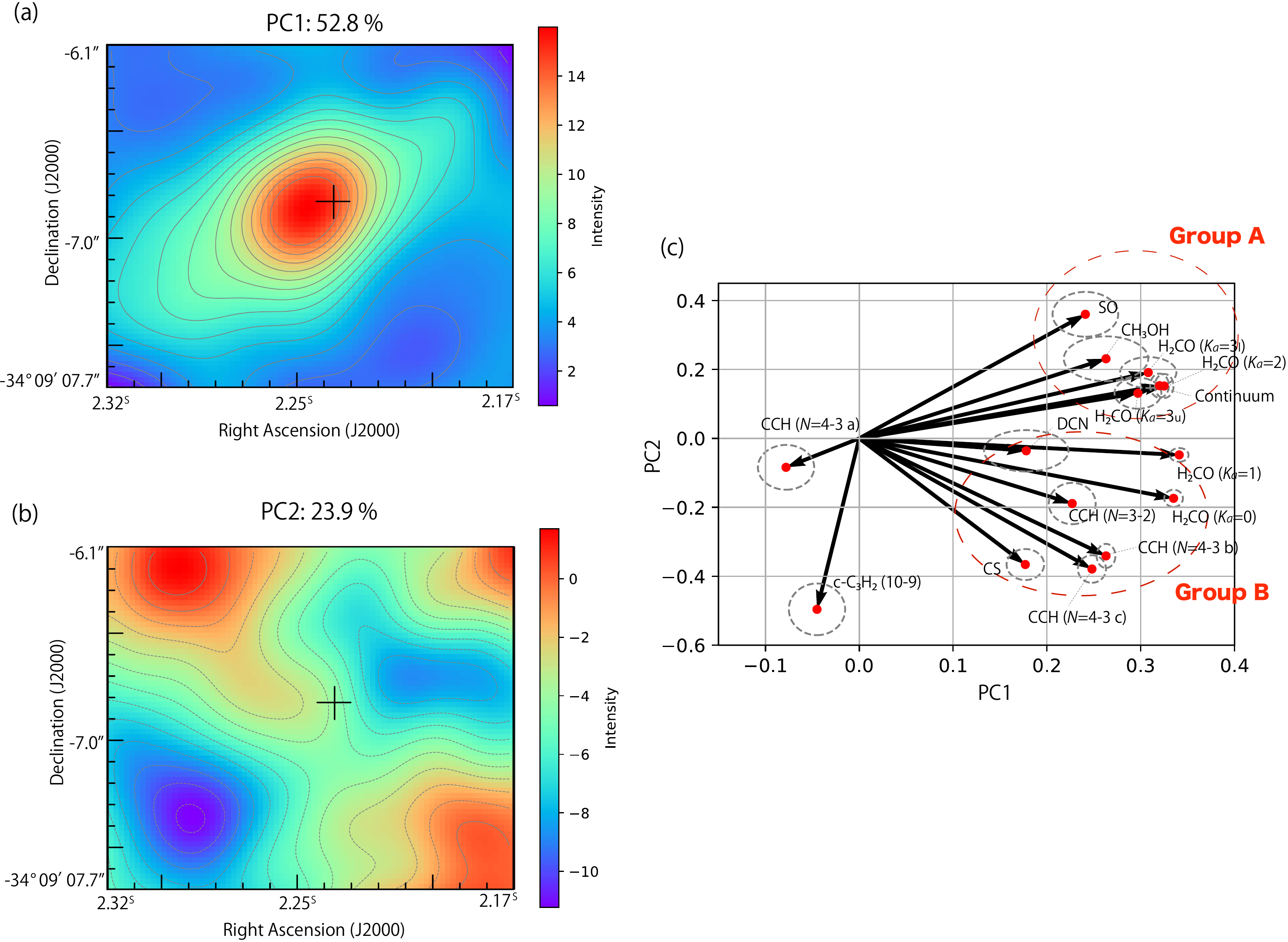}
\caption{(a,b) The principal components PC1 and PC2 for the disk/envelope structure (Figure \ref{protostar}). (c) The plot of the principal components for each distribution on the PC1-PC2 plane.  The cross marks show the continuum peak position. The contours are drawn with the solid and dotted lines, which mean the positive and negative levels, respectively. The contour level interval is 1.0 starting from 0.0.
The red dashed circles show the groups of the compact distribution (Group A) and the extended distribution (Group B).
The grey dashed ellipses represent the uncertainties (See Section 4.3).
\ccchh\ ($10-9$), CCH ($N=3-2$), ($N=4-3$ a), ($N=4-3$ b), ($N=4-3$ c), \hhco\ ($K_a$=0), ($K_a$=1), ($K_a$=2), ($K_a$=3${_u}$), and ($K_a$=3${_l}$) denote \ccchh\ (10$_{0,10}-9_{1,9}$ and $10_{1,10}-9_{0,9}$), CCH ($N=3-2, J=7/2-5/2, F=4-3$ and $3-2$), ($N=4-3, J=7/2-5/2, F=3-3$), ($N=4-3, J=7/2-5/2, F=4-3$ and $ 3-2$), ($N=4-3, J=9/2-7/2, F=5-4$ and $4-3$), \hhco\ (5$_{0,5}-4_{0,4}$), (5$_{1,5}-4_{1,4}$), (5$_{2,4}-4_{2,3}$), (5$_{3,2}-4_{3,1}$), and (5$_{3,3}-4_{3,2}$), respectively.\label{pc_protostar}}
\end{figure}

\begin{figure}[h!]
\centering
\includegraphics[scale=0.55]{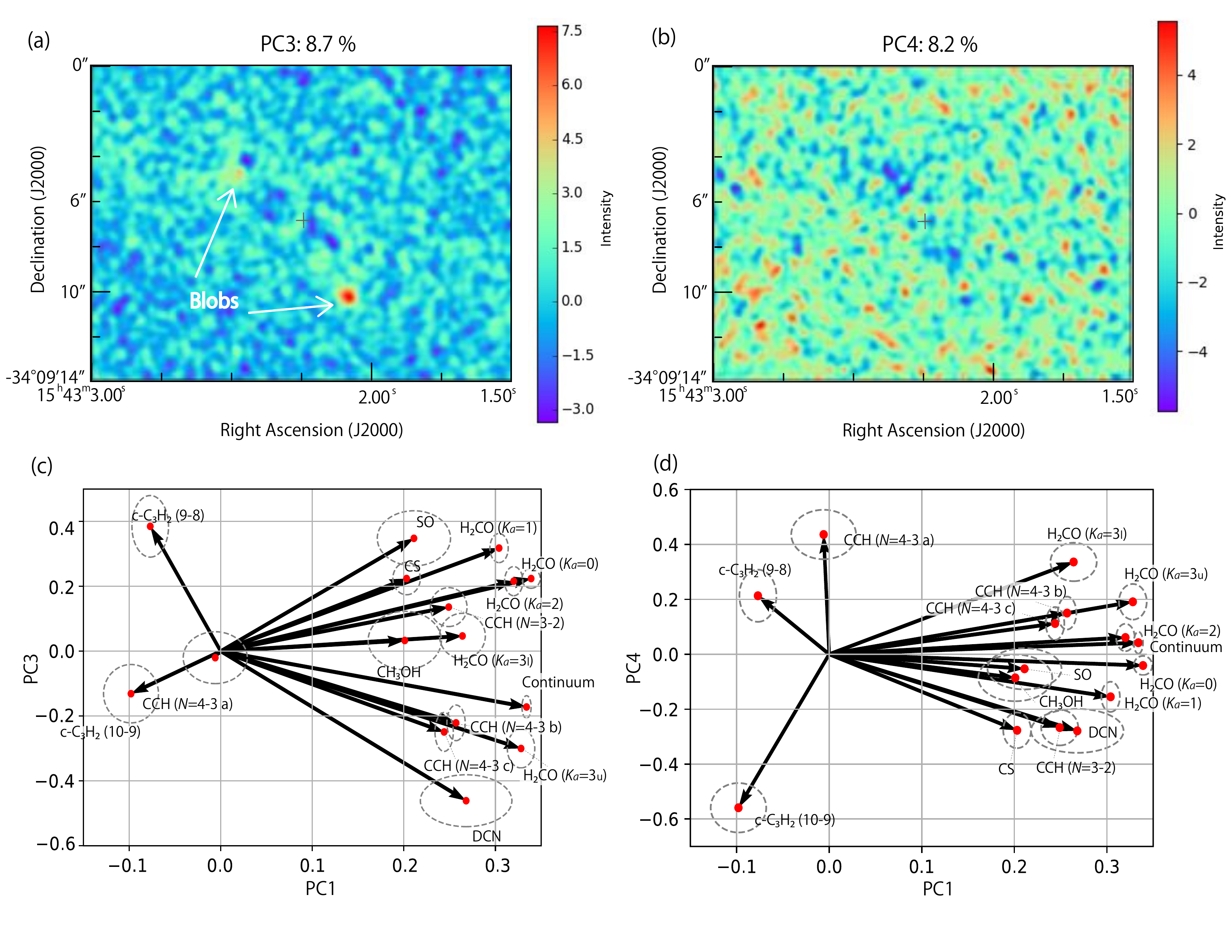}
\caption{(a,b) The principal components PC3 and PC4 for the whole-scale structure (Figure \ref{outflow}). (c,d) The plot of the principal components for each distribution on the PC1-PC3 and PC1-PC4 planes.
The cross marks show the continuum peak position. 
The grey dashed ellipses represent the uncertainties (See Section 4.3).
Blobs in (a) is consistent with the blobs A and D in Figure \ref{temperature_region}.
\ccchh\ ($10-9$), ($9-8$), CCH ($N=3-2$), ($N=4-3$ a), ($N=4-3$ b), ($N=4-3$ c), \hhco\ ($K_a$=0), ($K_a$=1), ($K_a$=2), ($K_a$=3${_u}$), and ($K_a$=3${_l}$) denote \ccchh\ (10$_{0,10}-9_{1,9}$ and $10_{1,10}-9_{0,9}$), \ccchh\ (9$_{1,8}-8_{2,7}$ and 9$_{2,8}-8_{1,7}$), CCH ($N=3-2, J=7/2-5/2, F=4-3$ and $3-2$), ($N=4-3, J=7/2-5/2, F=3-3$), ($N=4-3, J=7/2-5/2, F=4-3$ and $ 3-2$), ($N=4-3, J=9/2-7/2, F=5-4$ and $4-3$), \hhco\ (5$_{0,5}-4_{0,4}$), (5$_{1,5}-4_{1,4}$), (5$_{2,4}-4_{2,3}$), (5$_{3,2}-4_{3,1}$), and (5$_{3,3}-4_{3,2}$), respectively. \label{outflow_pc3_4}}
\end{figure}

\begin{figure}[h!]
\centering
\includegraphics[scale=0.55]{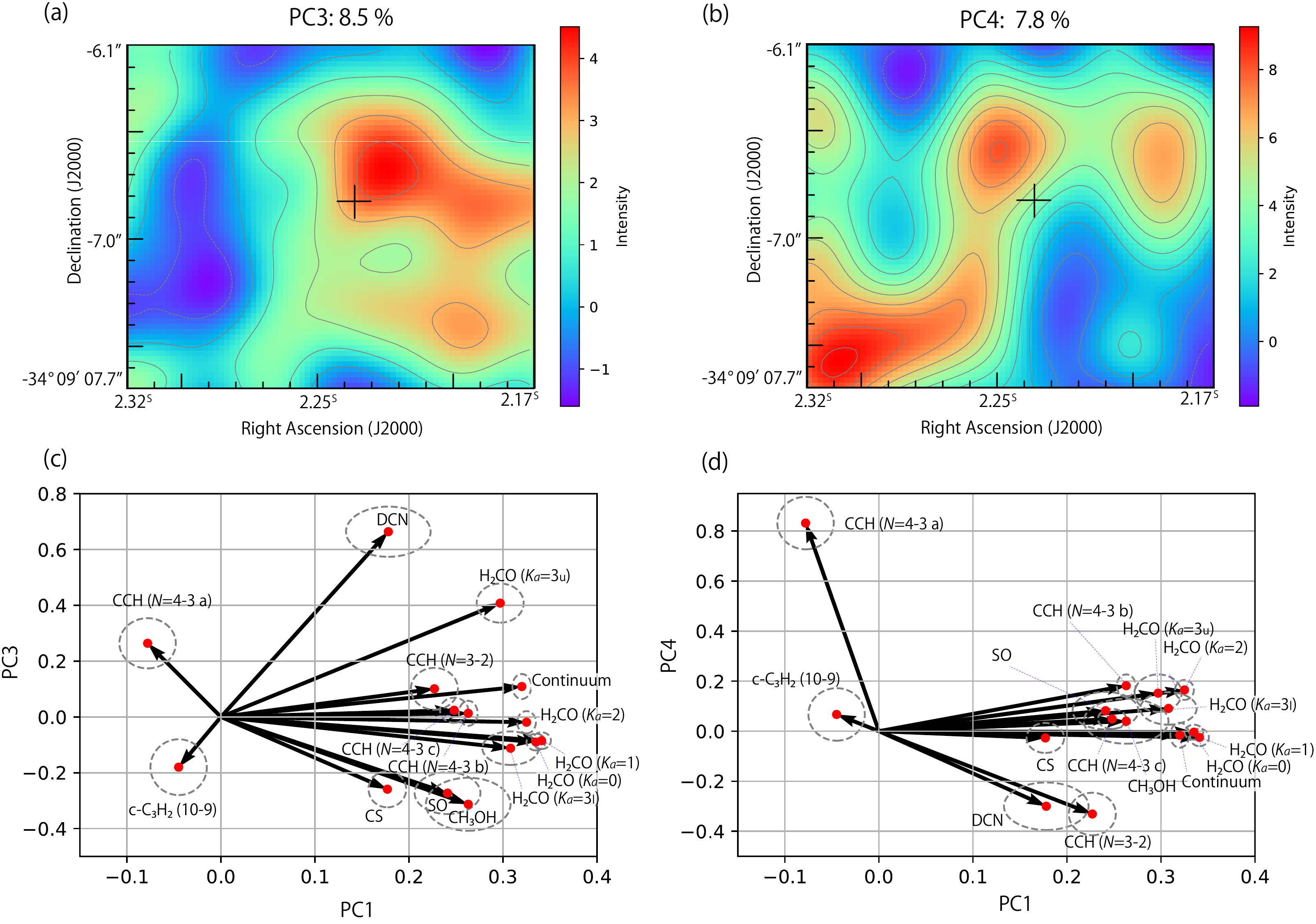}
\caption{(a,b) The principal components PC3 and PC4 for the disk/envelope structure (Figure \ref{protostar}). (c,d) The plot of the principal components for each distribution on the PC1-PC3 and PC1-PC4 planes.
The cross marks show the continuum peak position. The contours are drawn with the solid and dotted lines, which mean the positive and negative levels, respectively. The contour level interval is 1.0 starting from 0.0.
The grey dashed ellipses represent the uncertainties (See Section 4.3).
\ccchh\ ($10-9$), CCH ($N=3-2$), ($N=4-3$ a), ($N=4-3$ b), ($N=4-3$ c), \hhco\ ($K_a$=0), ($K_a$=1), ($K_a$=2), ($K_a$=3${_u}$), and ($K_a$=3${_l}$) denote \ccchh\ (10$_{0,10}-9_{1,9}$ and $10_{1,10}-9_{0,9}$), CCH ($N=3-2, J=7/2-5/2, F=4-3$ and $3-2$), ($N=4-3, J=7/2-5/2, F=3-3$), ($N=4-3, J=7/2-5/2, F=4-3$ and $ 3-2$), ($N=4-3, J=9/2-7/2, F=5-4$ and $4-3$), \hhco\ (5$_{0,5}-4_{0,4}$), (5$_{1,5}-4_{1,4}$), (5$_{2,4}-4_{2,3}$), (5$_{3,2}-4_{3,1}$), and (5$_{3,3}-4_{3,2}$), respectively.\label{protostar_pc3_4}}
\end{figure}

\begin{figure}[h!]
\centering
\includegraphics[scale=0.55, angle=90]{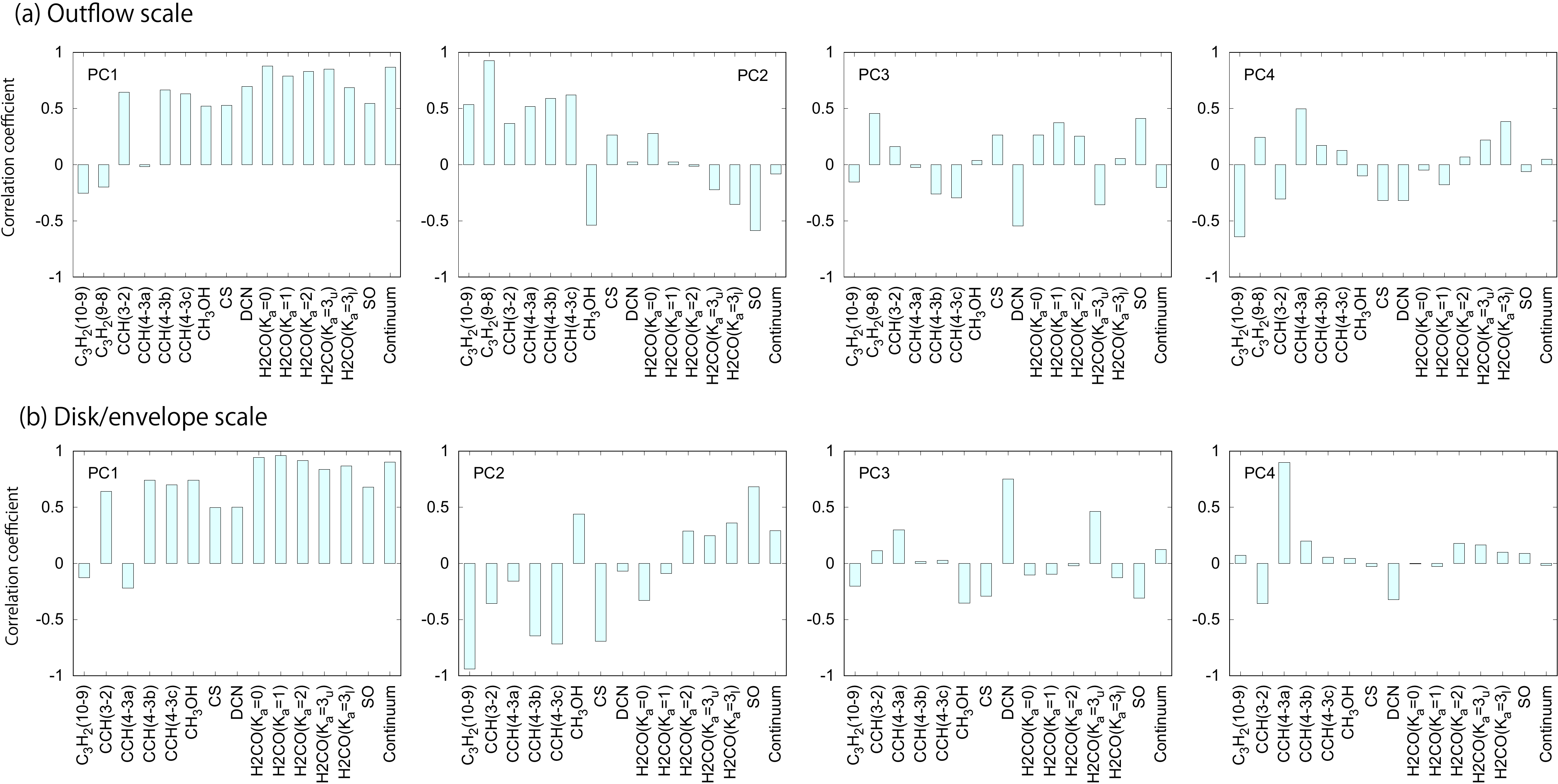}
\caption{(a,b) Correlation coefficients between the first four principal components and the molecular distributions of the whole-structure and disk/envelope scales, respectively. \label{load}}
\end{figure}



\end{document}